\documentclass[prb,preprint]{revtex4}

\usepackage{graphicx}
\usepackage{amsmath}
\usepackage{array,arydshln}
\usepackage{float}
\begin{document}

\title{Microwave non-reciprocity of magnon excitations in a non-centrosymmetric antiferromagnet Ba$_2$MnGe$_2$O$_7$}

\author{Y. Iguchi$^1$, Y. Nii$^1$, M. Kawano$^1$, H. Murakawa$^2$, N. Hanasaki$^2$, and Y. Onose$^1$} 
\affiliation{
$^1$Department of Basic Science, University of Tokyo, Tokyo, 153-8902, Japan\\
$^2$Department of Physics, Osaka University, Osaka, 560-0043, Japan}


\begin{abstract}
\bf{
We have investigated the microwave non-reciprocity for a non-centrosymmetric antiferromagnet Ba$_2$MnGe$_2$O$_7$. The magnon modes expected by the conventional spin wave theory for staggered antiferromagnets are certainly observed. The magnitudes of exchange interaction and magnetic anisotropy are obtained by the comparison with the theory. The microwave non-reciprocity is identified for one of these mode. The relative magnitude of microwave non-reciprocity can be explained with use of  spin wave theory and Kubo formula.
}
\end{abstract}
\maketitle

\section{Introduction}
Simultaneous breaking of spatial inversion and time reversal symmetries give rise to unique material properties. For example, the electric polarization is induced by the magnetic field, and reciprocally, the magnetization by the electric field in the symmetries-broken systems, which is termed magnetoelectric (ME) effect\cite{Dzyaloshinskii,Astrov,Folen}. The giant ME effect has been observed in many multiferroic materials\cite{Nature2006,Tokura2014}. The high frequency ME response can induce the unique properties of electromagnetic wave in media; the reflective index for wave vector $+\textbf{k}$ becomes different from that for $-\textbf{k}$, which is denoted as non-reciprocal directional dichroism (NDD) or birefringence. Rikken \textit{et al}., first observed the NDD in a chiral molecule under a magnetic field\cite{Rikken1997}. Similar NDD has been discerned in many multiferroic materials in the optical and X-ray regions\cite{Kubota2004,Jung2004,Saito2008}. Recently NDD has been reported also in terahertz(0.1-10 THz)\cite{Kezsmarki2011,Takahashi2011,Bordacs2012,Takahashi2013,Kibayashi2014,Kezsmarki2014,Kezsmarki2015,Bordacs2015,Takahashi2016,Narita2016} and microwave(1-100 GHz)\cite{Okamura2013,Tomita2014,Okamura2015,Nii2017,Iguchi2017} regions. The characteristic of these frequency regions is that the excitations in  magnetic materials are mostly caused by magnetic resonances. NDD for the magnetic resonances in ferromagnetic and helimagnetic states has been studied, extensively. NDD for antiferromagnetic resonance has also been studied with use of terahertz technique in a non-centrosymmetric antiferromagnet Ba$_2$CoGe$_2$O$_7$\cite{Kezsmarki2011,Bordacs2012,Kezsmarki2014}. While large NDD was successfully observed for Ba$_2$CoGe$_2$O$_7$, the intra-atomic magnetic transition was overlapped and the low energy antiferromagnetic resonance was observed only in the high magnetic field region. In order to extensively analyze the NDD for the antiferromagnetic resonance with use of simple spin wave theoretical model, we have investigated a related material Ba$_2$MnGe$_2$O$_7$.


Ba$_2$MnGe$_2$O$_7$ has the same non-centrosymmetric crystal structure as Ba$_2$CoGe$_2$O$_7$[Fig. 1(a)], but the Mn$^{2+}$ ions replace the Co$^{2+}$ ions\cite{Zheludev2003,Masuda2010}. Mn$^{2+}$ ion has isotropic $S=5/2$ state because all the five \textit{d} orbitals are singly occupied. The staggered antiferromagnetic structure is realized below $T_N =$ 4 K\cite{Masuda2010}. In the in-plane magnetic field, the magnetic structure is rotated so that the staggered component of magnetic moment is perpendicular to the external magnetic field as shown in Fig. 1(b).   The magnetic exchange interaction between nearest neighboring Mn moments is $\approx 27 \mu$eV, which is smaller than that of Ba$_2$CoGe$_2$O$_7$ ($\approx$ 230 $\mu$eV)\cite{Masuda2010,Penc}. Therefore, the energy scale of antiferromagnetic resonance, which is determined by the geometric mean of  exchange interaction and magnetic anisotropy\cite{Gurevich}, is much lower than Ba$_2$CoGe$_2$O$_7$. Here we have successfully observed the antiferromagnetic magnon modes of Ba$_2$MnGe$_2$O$_7$ in the magnetic fields (0.1-5 T) with use of microwave technique. Moreover we have identified finite microwave non-reciprocity for one of the antiferromagnetic magnon modes. By using the ME coupling constant obtained by the static ME effect, we have found the observed NDD can be {\it quantitatively} explained by the spin wave theory and Kubo formula.

\begin{figure}
\begin{center}
\includegraphics*[width=15cm]{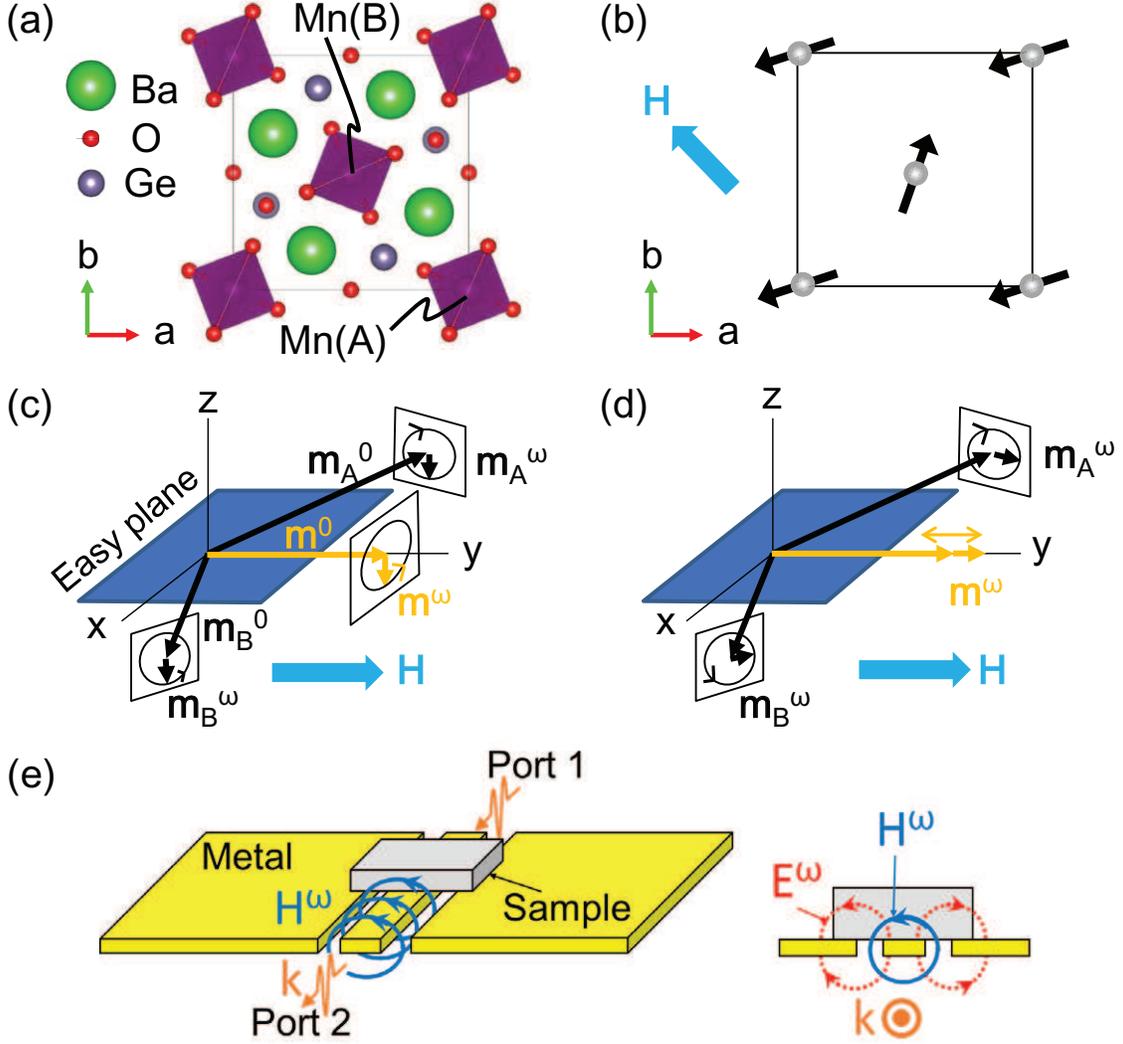}
\caption{
(a) Crystal structure of Ba$_2$MnGe$_2$O$_7$. The two Mn ions are denoted as Mn(A) and Mn(B).
(b) Magnetic structure in the magnetic field along the [$\bar{1}10$] direction of Ba$_2$MnGe$_2$O$_7$.
(c),(d) Illustrations of two magnon modes ((c) mode 1, (d) mode 2) in the in-plane magnetic field for an easy-plane antiferromagnet. $\textbf{m}_\mathrm{A}^0$, $\textbf{m}_\mathrm{A}^\omega$, $\textbf{m}_\mathrm{B}^0$, and $\textbf{m}_\mathrm{B}^\omega$ are the static and dynamical magnetic moments at the sublattice A and sublattice B, respectively. $\textbf{m}^0$ and $\textbf{m}^\omega$ are the static and dynamical parts of total magnetic moment, respectively. 
(e) Sketch of experimental setup. The sample is put at the center of coplanar wave guide. The microwave propagates along the center signal line. The alternating magnetic field $\textbf{H}^\omega$ and the alternating electric field $\textbf{E}^\omega$ are perpendicular to the microwave wave vector $\textbf{k}$.}
\end{center}
\end{figure}

\section{Experimental method}



We prepared single crystals of Ba$_2$MnGe$_2$O$_7$ by using the Floating zone method\cite{Murakawa2012}. 
We measured the microwave absorption on the coplanar waveguide, which  was designed so that the characteristic impedance coincides 50 $\mathrm{\Omega}$. The width of the signal line was 0.2 mm, and the gap between the signal line and ground planes was 0.05 mm. The single crystal was put on the center of waveguide and measured the microwave absorption in the external magnetic field ($\textbf{H}$). The microwave absorption spectra $\Delta S_{12}$ was deduced by the difference of $S_{12} (H)$ from the zero field value. Here, $S_{12}$ is the transmittance coefficient from port 2 to port 1 (The two ports are connected to the two terminals of waveguide). In this case, we used the zero field data as the background because the present antiferromagnetic samples show negligible microwave absorption at $H = 0$. $\Delta S_{21}$ is the absorption of microwave for the wave vector opposite to the case of $\Delta S_{12}$. The alternating magnetic field of microwave ($\textbf{H}^\omega$) is induced in the plane perpendicular to the wave vector $\textbf{k}$. Hereafter, we specify which crystal axes are along $\textbf{H}$ and perpendicular to $\textbf{H}^\omega$ in order to describe the experimental geometry. The microwave absorption was measured in a superconducting magnet with use of a vector network analyzer (N5230A, Agilent). All the experimental data in this paper were taken at $T = 1.8$ K. 


\section{Results and discussions}
Figure 2(a) shows the microwave absorption spectra at various magnetic fields for $\textbf{H}\parallel[1\bar{1}0]$ and $\textbf{H}^\omega\perp[110]$. We have identified two peaks in the absorption spectra. One peak is observed in the low frequency region at a low magnetic field. The peak frequency and intensity increase with the magnetic field. This mode is denoted as mode 1. The other mode is observed around 26 GHz in the low field region. The peak frequency is almost unchanged below 1 T but gradually decreases with the magnetic field above 1 T. This magnon mode is denoted as mode 2. 
\begin{figure}
\begin{center}
\includegraphics*[width=15cm]{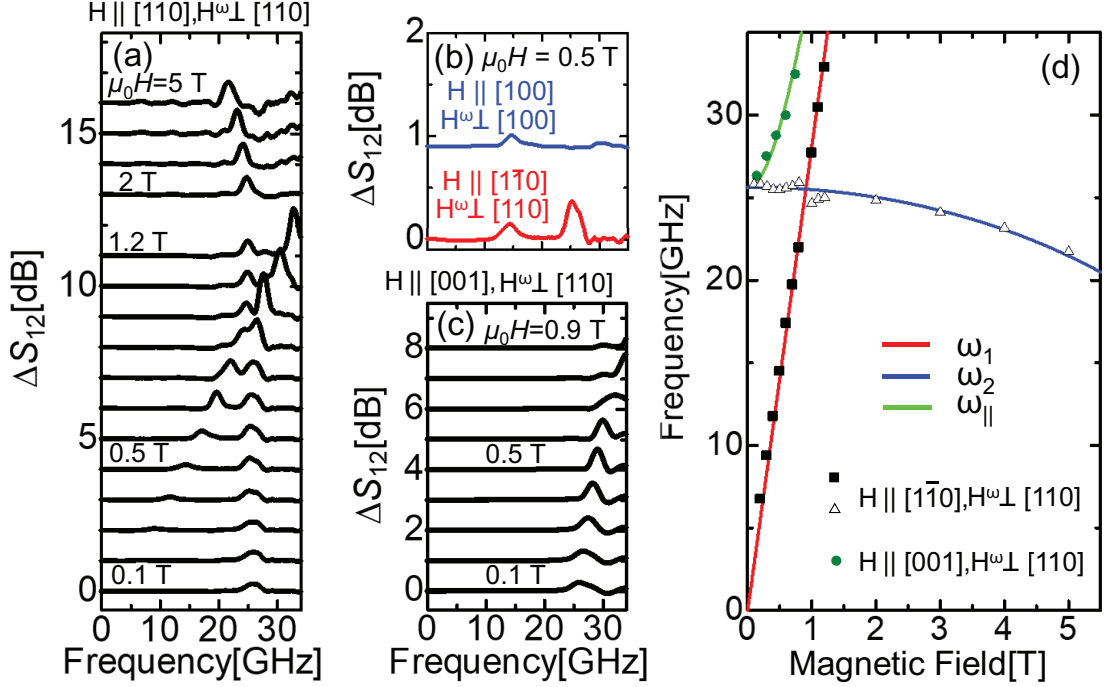}
\caption{
(a) The microwave absorption spectra $\Delta S_{12}$ at various magnetic fields in the experimental geometry with $\textbf{H}\parallel[1\bar{1}0]$ and $\textbf{H}^\omega \perp [110]$.
(b) Comparison of $\Delta S_{12}$ at 0.5 T in the two geometries, $\textbf{H}\parallel[100], \textbf{H}^\omega\perp[100]$ and $\textbf{H}\parallel[1\bar{1}0]$, $\textbf{H}^\omega\perp[110]$.
(c) $\Delta S_{12}$ at various magnetic fields for $\textbf{H}\parallel[001]$ and $\textbf{H}^\omega \perp [110]$.
(d) Experimentally observed and theoretically obtained magnon frequencies. Squares and triangles indicate the peak frequencies of experimentally observed magnon mode 1 and mode 2 for $\textbf{H}\parallel[1\bar{1}0]$, $\textbf{H}^\omega \perp [110]$, respectively, and circles the experimentally observed magnon mode for $\textbf{H}\parallel[001]$, $\textbf{H}^\omega \perp [110]$. The corresponding theoretical curves [Eqs. (1)-(3)] are plotted as solid lines.}
\end{center}
\end{figure}
The peak frequencies are plotted as a function of magnetic field in Fig. 2(d). While the frequency of mode 1 increases linearly with the magnetic field, that of mode 2 gradually decreases as the magnetic field is increased. To examine the origin of these magnon modes, we measured the polarization dependence of absorption spectra. We have found that the absorption peak for mode 2 is absent for $\textbf{H}\parallel[100]$ and $\textbf{H}^\omega\perp[100]$ as shown in Fig. 2(b). This indicates the alternating magnetization in mode 2 is along the external static magnetic field. Actually such a polarization dependence is expected for the conventional magnon modes in easy-plane antiferromagnet in the in-plane magnetic field.  Figures 1(c) and 1(d) illustrate the conventional magnon modes. The one mode is uniform oscillation of magnetic moment with keeping the relative angle of magnetic moments [Fig. 1(c)]. The other mode is the anti-phase oscillation of two magnetic moments in a unit cell [Fig. 1(d)]. The oscillation of total magnetic moment is along the external magnetic field. The mode 2 seems to correspond to the latter magnon modes judging from the polarization dependence while the mode 1 seems the former magnon mode. Theoretically, the frequencies of mode 1 and mode 2 are expressed as
\begin{eqnarray}
\omega_{1} & = & \gamma \mu_0 H \sqrt{1 + \frac{H_\mathrm{A}}{2H_\mathrm{E}}}, \\
\omega_{2} & = &  \gamma \mu_0 \sqrt{ 2H_\mathrm{E} H_\mathrm{A} - \frac{H_\mathrm{A}}{2H_\mathrm{E}}H^2}.
\label{eq:omega12}
\end{eqnarray}
Here $\gamma,\mu_0, H_\mathrm{A}$, and $H_\mathrm{E}$ are the gyromagnetic ratio, the magnetic permeability in vacuum, the magnetic anisotropy field, and the exchange field, respectively. As shown in Fig. 2(d), these theoretical formula are quite consistent with the experimental observation. To further examine the theory-experiment correspondence, we study the magnon in the magnetic field along [001] direction. In this case, one mode is zero frequency rotation of magnetic moments around the [001] direction. Therefore, only one mode is expected in the finite frequency  regime. We certainly observe only one magnon peak in this experimental geometry [Fig. 2(c)]. The magnetic field dependence of frequency is theoretically expressed as follow\cite{Gurevich};
\begin{equation}
\omega_{\parallel} = \gamma \mu_0 \sqrt{ 2 H_\mathrm{E} \frac{2H_\mathrm{E} + H_\mathrm{A}}{(2H_\mathrm{E} - H_\mathrm{A})^2} H^2 + 2H_\mathrm{E} H_\mathrm{A} }.
\label{eq:omegapara}
\end{equation}
The experimental data of peak frequency is reproduced with the same parameters as the in-plane-field case. From the fittings of experimental data to the theoretical formula, we obtained $\mu_0H_A \simeq 0.09$ T and $\mu_0H_E \simeq 4.67$ T, which are corresponding to the exchange interaction constant $J \simeq 27$ $\mu$eV and the single ion anisotropy $K \simeq 2$ $\mu$eV, respectively. While the estimated exchange interaction almost coincides with that estimated by the previous neutron scattering study\cite{Masuda2010}, the magnitude of magnetic anisotropy in this system was not reported previously. Reflecting the isotropic $S=5/2$ state, the magnetic anisotropy is much smaller than the isostructural Ba$_2$CoGe$_2$O$_7$ (1.4 meV)\cite{Miyahara2011,Penc}.

As mentioned above, microwaves are expected to show the non-reciprocity in time reversal and spatial inversion symmetries simultaneously broken systems. We tried to observe the microwave non-reciprocity in two experimental geometries. The first geometry is $\textbf{H}\parallel \left\langle 100\right\rangle $, $\textbf{H}^\omega \perp \textbf{H}$. In this case, only the mode 1 is observable. In the magnetic field along [100], the magnetic symmetry is chiral\cite{Bordacs2012} and expected to show the non-reciprocity for counter-propagating microwave along the magnetic field direction. We show the microwave absorption spectra $\Delta S_{12}$ and $\Delta S_{21}$ at 0.4 T for $\textbf{H}\parallel[100]$ in Fig. 3(a). We have found that $\Delta S_{12}$ and $\Delta S_{21}$ are different from each other. The difference of absorptions $\Delta S_{12} - \Delta S_{21}$ indicates the microwave non-reciprocity. It was reversed in the reversal magnetic field as shown in Fig. 3(b). It should be noted that the 90 degree rotation of sample around the [001] direction corresponding to the spatial inversion operation, and the chirality and microwave non-reciprocity should be reversed when $\textbf{H} \parallel \textbf{k} \parallel [010]$\cite{Bordacs2012}. In order to discuss the effect of spatial inversion on the microwave non-reciprocity, we show the microwave non-reciprocity for $\textbf{H} \parallel \textbf{k} \parallel [010]$. As shown in Figs. 3(c) and 3(d), the microwave non-reciprocity is reversed by the spatial inversion. When the magnetic field is increased, the magnitude of non-reciprocity increases as shown in Fig. 3(e). 

\begin{figure}
\begin{center}
\includegraphics*[width=16cm]{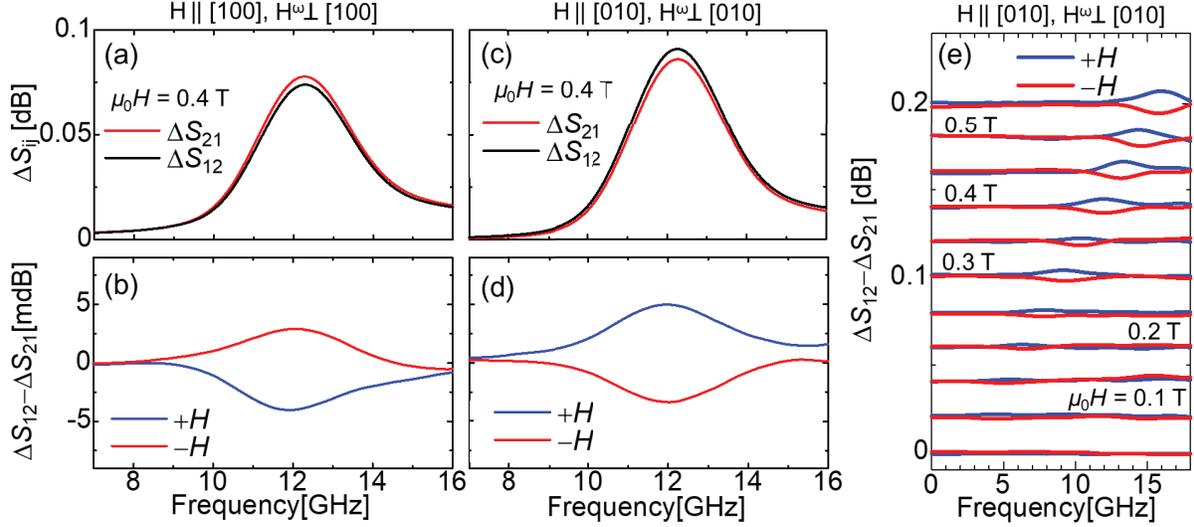}
\caption{
(a) Microwave absorption spectra $\Delta S_{12}$ and $\Delta S_{21}$ for $\textbf{H}\parallel[100]$ and $\textbf{H}^\omega\perp [100]$. (b) Microwave non-reciprocity $\Delta S_{12} - \Delta S_{21}$ at $\pm$0.4 T for $\textbf{H}\parallel[100]$ and $\textbf{H}^\omega\perp [100]$.
(c) $\Delta S_{12}$ and $\Delta S_{21}$ for $\textbf{H}\parallel[010]$ and $\textbf{H}^\omega\perp [010]$. (d) $\Delta S_{12} - \Delta S_{21}$ at $\pm$0.4 T for $\textbf{H}\parallel[010]$ and $\textbf{H}^\omega\perp [010]$. (e) $\Delta S_{12} - \Delta S_{21}$ for $\textbf{H}\parallel[010]$ and $\textbf{H}^\omega\perp [010]$ at various positive and negative magnetic fields.
}
\end{center}
\end{figure}


Let us move on to the second geometry of microwave non-reciprocity measurement, where $\textbf{H}\parallel[1\bar{1}0]$ and $\textbf{H}^\omega \perp [110]$. In this case, the sample has an electric polarization along [001], and both the mode 1 and the mode 2 are observable. Figures 4(a)-(c) and 4(d)-(f) show the microwave absorption spectra around the mode 1 and the mode 2, respectively. One can see that the non-reciprocities in this low magnetic field are almost negligible in this experimental geometry. It should be noted that the non-reciprocity caused by the magnetic dipolar interaction\cite{Nii2017}, which is distinct from the non-reciprocity due to the material symmetry breaking, becomes dominant in the high magnetic field region above 0.5 T. The dipolar non-reciprocity was not reversed by the 90 degree rotation of sample around the [001] direction, which is equivalent to the spatial inversion.

\begin{figure}
\begin{center}
\includegraphics*[width=14cm]{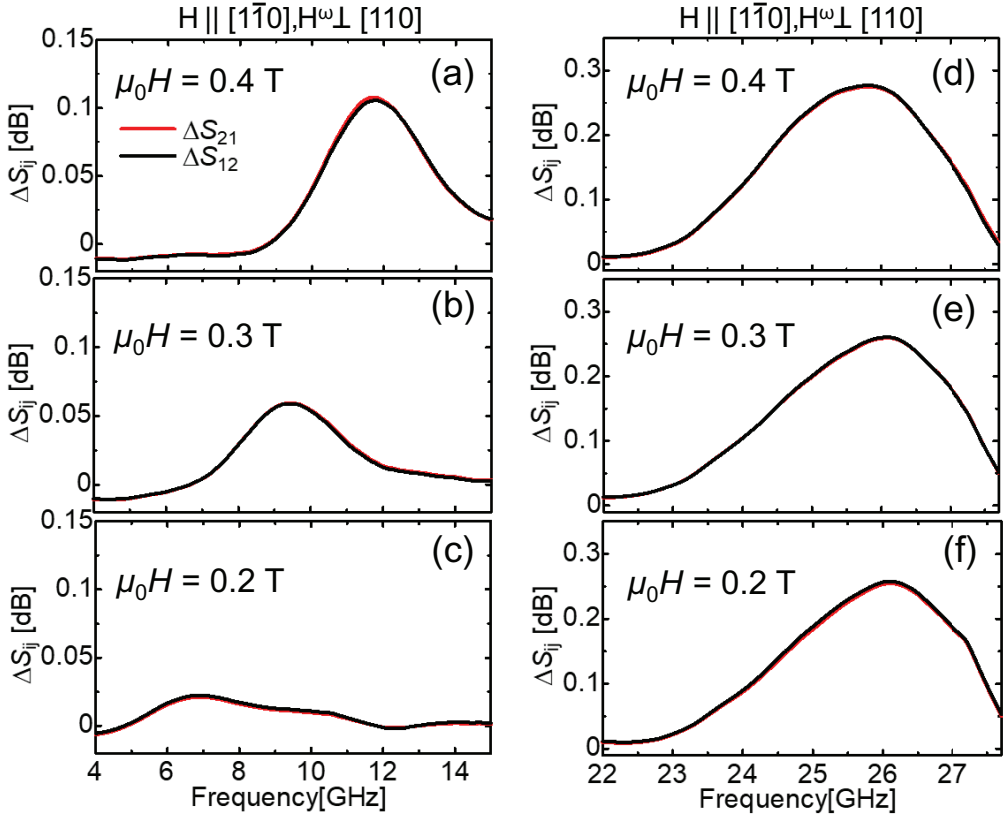}
\caption{
(a)-(f)Microwave absorption spectra $\Delta S_{12}$ and $\Delta S_{21}$ for $\textbf{H}\parallel[1\bar{1}0]$ and $\textbf{H}^\omega \perp [110]$. (a), (b), and (c) show the spectra around the frequency of mode 1 at 0.4 T, 0.3 T, and 0.2 T, respectively. (d), (e), and (f) around the frequency of mode 2 at 0.4 T, 0.3 T, and 0.2 T, respectively. 
}
\end{center}
\end{figure}

Finally, let us compare the observed microwave non-reciprocity with the theoretical calculation. Theoretically, the relative non-reciprocity for the linearly polarized microwave with $\textbf{k} \parallel \textbf{x}$ and $\textbf{H}^\omega \parallel \textbf{z}$, and $\textbf{E}^\omega \parallel \textbf{y}$ can be expressed as\cite{supple} 
\begin{equation}
\frac{\Delta S_{12} - \Delta S_{21}}{\Delta S_{12} + \Delta S_{21}} = \frac{\mathrm{Im}\left[ \chi^{me}_{zy}+\chi^{em}_{yz} \right] }{2\mathrm{Im}\left[ \sqrt{\left( 1 +  \chi^{mm}_{zz} \right) \left( \varepsilon_\infty + \chi^{ee}_{yy}\right) } \right] }, 
\label{eq:NDD}
\end{equation}
where $\chi^{me}_{ij}$, $\chi^{em}_{ij}$, $\chi^{ee}_{ij}$, $\chi^{mm}_{ij}$, $\varepsilon_\infty$ are magnetoelectric, electromagnetic, electric, and magnetic susceptibility tensors and high frequency relative dielectric constant, respectively.  
According to the Kubo formula, these susceptibilities are obtained by the following relations\cite{supple};
\begin{equation}
\chi_{ij}^{me} = \frac{NV}{\hbar}\sqrt{\frac{\mu_0}{\varepsilon_0}}\sum_n\frac{\left\langle 0 \left| \Delta M_i \right| n  \right\rangle \left\langle n \left| \Delta P_j \right| 0  \right\rangle }{\omega - \omega_n + i\delta},
\label{eq:chime}
\end{equation}
\begin{equation}
\chi_{ij}^{em} = \frac{NV}{\hbar}\sqrt{\frac{\mu_0}{\varepsilon_0}}\sum_n\frac{\left\langle 0 \left| \Delta P_i \right| n  \right\rangle \left\langle n \left| \Delta M_j \right| 0  \right\rangle }{\omega - \omega_n + i\delta},
\label{eq:chiem}
\end{equation}
\begin{equation}
\chi_{ij}^{mm} = \frac{NV}{\hbar}\mu_0\sum_n\frac{\left\langle 0 \left| \Delta M_i \right| n  \right\rangle \left\langle n \left| \Delta M_j \right| 0  \right\rangle }{\omega - \omega_n + i\delta},
\label{eq:chimm}
\end{equation}
\begin{equation}
\chi_{ij}^{ee} = \frac{NV}{\hbar}\frac{1}{\varepsilon_0}\sum_n\frac{\left\langle 0 \left| \Delta P_i \right| n  \right\rangle \left\langle n \left| \Delta P_j \right| 0  \right\rangle }{\omega - \omega_n + i\delta},
\label{eq:chiee}
\end{equation}
where $\Delta \textbf{M}$ and $\Delta \textbf{P}$ are, respectively, the dynamical polarization and magnetization induced by the magnon. The matrix element of $\Delta \textbf{M}$ can be deduced by using spin wave theory. For the calculation of $\Delta \textbf{P}$, we assume the \textit{d}-\textit{p} hybridization type magnetoelectric coupling and the coupling constant is determined by the fitting of dc magnetoelectric response measured by Murakawa \textit{et al}\cite{Murakawa2012}. For the detail of theoretical calculations, see the supplemental material\cite{supple}. In Fig. 5, we plot the theoretically calculated and experimentally observed relative non-reciprocity $(\Delta S_{12} - \Delta S_{21})/(\Delta S_{12} + \Delta S_{21})$ for the mode 1. Both the microwave absorption and the difference of $\Delta S_{12}$ and $\Delta S_{21}$ decrease with decreasing the magnetic field. The relative non-reciprocity gradually increases as the magnetic field is decreased. The theoretical calculation of relative non-reciprocity coincides with the experimental data with respect to both the magnitude and the field dependence. On the other hand, the theoretical value of non-reciprocity in the second experimental geometry is quite small compared with the first one, similarly to the experimental result. In this geometry, the static polarization shows a maximum as a function of angle of $\textbf{H}$\cite{Murakawa2012}, and the alternating electric polarization due to magnon excitation becomes quite small. For this reason, the non-reciprocity due to the dynamical ME effect is also quite small in this case. Thus, the microwave non-reciprocity in this system is quantitatively explained by the theoretical calculation, which give rise to the satisfactory understanding of microwave non-reciprocity in Ba$_2$MnGe$_2$O$_7$.

\begin{figure}
\begin{center}
\includegraphics*[width=12cm]{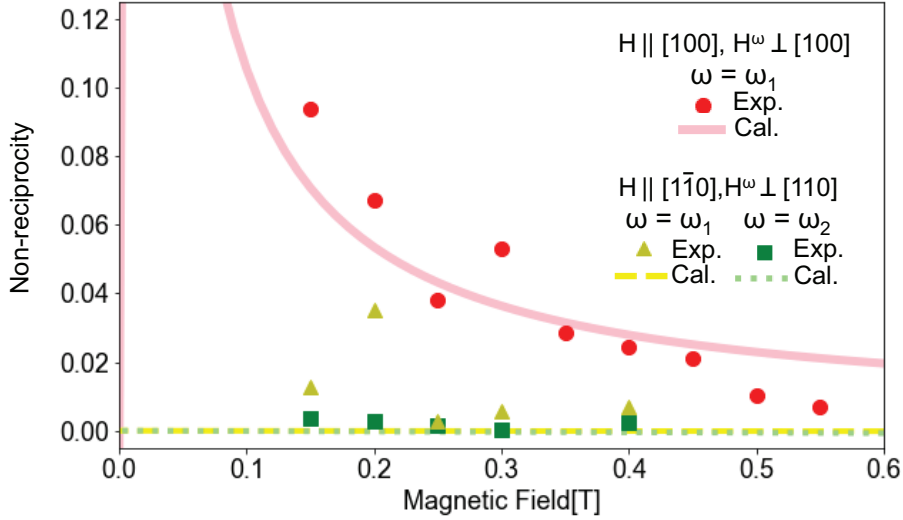}
\caption{
The relative microwave non-reciprocity $(\Delta S_{12} - \Delta S_{21})/(\Delta S_{12} + \Delta S_{21})$ for $\textbf{H}\parallel[100]$, $\textbf{H}^\omega \perp[100]$ for the mode 1 is plotted as circles. Solid line stands for the corresponding theoretical calculation. The relative microwave non-reciprocities for $\textbf{H}\parallel[1\bar{1}0]$, $\textbf{H}^\omega \perp[110]$ for mode 1 and mode 2 are plotted as triangles and squares, respectively. Dashed and dotted lines show the theoretical calculation of microwave non-reciprocity of mode 1 and mode 2 for $\textbf{H}\parallel[1\bar{1}0]$, $\textbf{H}^\omega \perp[110]$, respectively.
}
\end{center}
\end{figure}

\section{Summary}

In summary, we observed the antiferromagnetic magnon modes of Ba$_2$MnGe$_2$O$_7$ in the microwave region. The notable microwave non-reciprocity was observed for the mode 1 for $\textbf{H}\parallel[100]$ and $\textbf{H}^\omega\perp[100]$. On the other hand, it is negligible for both the mode 1 and mode 2 when $\textbf{H}\parallel[1\bar{1}0]$ and $\textbf{H}^\omega\perp[110]$. The presence /absence and magnitude of non-reciprocity are explained by the theoretical analysis based on the spin wave theory and Kubo formula. These quantitative experiment-theory correspondences adequately ensure the validity of background physics such as non-reciprocal microwave response and the \textit{d}-\textit{p} hybridization mechanism.

\section*{ACKNOWLEDGMENTS}
The authors thank {C. Hotta, S. Hirose, and K. Penc} for fruitful discussion. This work was in part supported by the Grant-in-Aid for Scientific Research (Grants Nos. {17H05176, 16H04008}) from the Japan Society for the Promotion of Science. Y.I. is supported by the Grant-in-Aid for Research Fellowship for Young Scientists from the Japan Society for the Promotion of Science (No. 16J10076).

\newpage

\begin{center}
	\Large
	{Supplemental Material for the article\lq\lq Microwave non-reciprocity of magnon excitations in a non-centrosymmetric antiferromagnet Ba$_2$MnGe$_2$O$_7$ \rq\rq} by Iguchi $et$ $al.$
\end{center}

\renewcommand{\theequation}{S\arabic{equation}}
\setcounter{equation}{0}
\renewcommand{\thefigure}{S\arabic{figure}}
\setcounter{figure}{0}

\section*{SI. Magnetic structure in magnetic fields}
In this supplemental material, we theoretically discuss the magnetic excitation and the microwave non-reciprocity in order to compare with the experimentally observed data. Similar calculations were already done in literatures\cite{Miyahara2011,Bordacs2012,Kezsmarki2014}. We assume the Hamiltonian in Ba$_2$MnGe$_2$O$_7$ is 
\begin{equation}
\mathcal{H} = J\sum_{<i,j>}\textbf{S}_i\cdot\textbf{S}_j + K\sum_{i}\left( S_i^z \right) ^2 + g\mu_\mathrm{B} \sum_{i}\textbf{S}_i\cdot\left( \mu_0\textbf{H} \right). 
\label{eq:Hamiltonian}
\end{equation}
Here, $g$ is a $g$ value, and $\mu_\mathrm{B}$ is the Bohr magneton. $\mu_0$ is the magnetic permeability in vacuum. $J$ is the nearest-neighbor exchange interaction constant. The nearest-neighbor exchange interaction is antiferromagnetic ($J>0$). The interplane magnetic interaction is small compared with the intraplane one\cite{Masuda2010}, therefore ignored here for simplicity. Dzyaloshinskii-Moriya interaction is also ignored. The single-ion anisotropy $K>0$ indicates the easy-plane-type magnetic anisotropy. $\textbf{S}_i = \left(S_i^x, S_i^y, S_i^z \right) $ is the spin operator at $i$ sublattice ($i = \mathrm{A}, \mathrm{B}$), the magnetic moment is $\textbf{m}_i = -g\mu_\mathrm{B}\textbf{S}_i$.  

\begin{figure}[tbh]
	\centering
	\includegraphics[width=0.4\linewidth]{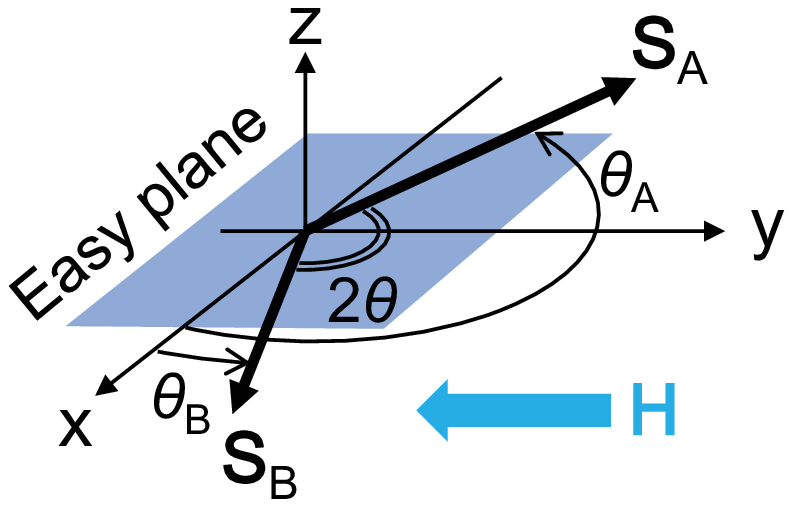}
	\caption{
		The ground state of easy-plane-type antiferromagnets in the inplane magnetic field $\textbf{H}$. 
	}
	\label{fig:AFM}
\end{figure}

In this section, we deduce the magnetic structure in magnetic fields at $T = 0$ K with use of classical approach. We assume two-sublattice magnetic structure. The magnetic field is applied in the tetragonal plane. Therefore, the magnetic field vector can be expressed as
\begin{equation}
\textbf{H} = H \left( \cos\theta_\mathrm{H} , \sin\theta_\mathrm{H}, 0 \right). 
\label{eq:Ham2}
\end{equation}
In this case, the spins for each sublattice are vector along the tetragonal plane expressed as  
\begin{equation}
\textbf{S}_i = S\left( \cos\theta_i, \sin\theta_i,0 \right),
\label{eq:classicspin}
\end{equation}
where $\theta_i (i=\mathrm{A,B})$ stands for the angle of spin for the $i$ sublattice. Then the energy is estimated as
\begin{equation}
\frac{E}{N} = 4JS^2 \cos 2\theta + hS\left\lbrace \cos\left( \theta_\mathrm{A} - \theta_\mathrm{H} \right) + \cos\left(\theta_\mathrm{B} - \theta_\mathrm{H} \right)   \right\rbrace,  
\label{eq:E/N}
\end{equation}
where $2\theta = \theta_\mathrm{A} - \theta_\mathrm{B}\hspace{0.5cm}  \left( \theta_\mathrm{A} > \theta_\mathrm{B} \right),\hspace{0.5cm}  h = g\mu_\mathrm{B} \mu_0 H$. $N$ is the number of unit cell. Neglecting the finite temperature effect, the spins are ordered so that the energy is minimized. From the condition, we obtain the directions of spins as follows;
\begin{equation}
\cos\theta = \frac{h}{8JS},
\label{eq:costheta}
\end{equation}
\begin{equation}
\theta_\mathrm{A} = \theta_\mathrm{H} + \theta + \pi,\hspace{0.5cm}  \theta_\mathrm{B} = \theta_\mathrm{H} - \theta +\pi. 
\label{eq:costheta2}
\end{equation}
The obtained magnetic structure is shown in Fig. \ref{fig:AFM}(a).

\section*{SII. Electric polarization}
The electric polarization of Ba$_2$MnGe$_2$O$_7$ can be induced by the metal ligand hybridization mechanism. The local electric dipole moment at $i$ sublattice is described as
\begin{eqnarray}
\textbf{p}_{i} &=& \lambda \sum_{j} (\textbf{S}_{i} \cdot \textbf{e}_{ij})^2 \textbf{e}_{ij},
\label{eq:DeltaP}
\end{eqnarray}
where $\lambda$ is a constant and $\textbf{e}_{ij} = (e_{ij}^x, e_{ij}^y, e_{ij}^z)$ is the unit vector along the bond connecting Mn ion at $i$ sublattice and $j$th coordinated oxygen ion. For Ba$_2$MnGe$_2$O$_7$, the lattice constants are $a = b =$ 8.5022 \AA\hspace{0.1cm} and $c =$ 5.5244 \AA. In the unit cell, the Mn ions are located at the positions (0,0,0) and $(0.5a, 0.5a,0)$. The four coordinated oxygens around the Mn A ion are at ($0.0825a, 0.187a, 0.2117c$), ($-0.0825a, -0.187a, 0.2117c$), ($-0.187a, 0.0825a, -0.2117c$), and ($0.187a, -0.0825a, -0.2117c$). On the other hand, The coordinated oxygens around Mn B are at ($0.687a, 0.5825a, 0.2117c$), ($0.313a, 0.4175a, 0.2117c$), ($0.4175a, 0.687a, -0.2117c$), ($0.5825a, 0.313a, -0.2117c$). From these informations, we obtained
\begin{eqnarray}
\textbf{e}_{A1}& = &(0.33487, 0.75903, 0.55833)= (d,f,l),  \nonumber \\
\textbf{e}_{A2}& = &(-d, -f, l),  \nonumber \\
\textbf{e}_{A3}& = &(-f, d, -l),  \nonumber \\
\textbf{e}_{A4}& = &(f, -d, -l),  \nonumber
\label{eq:e1j}
\end{eqnarray}
\begin{eqnarray}
\textbf{e}_{B1}& = &(f, d, l),  \nonumber \\
\textbf{e}_{B2}& = &(-f, -d, l),  \nonumber \\
\textbf{e}_{B3}& = &(-d, f, -l),  \nonumber \\
\textbf{e}_{B4}& = &(d, -f, -l).  \nonumber
\label{eq:e2j}
\end{eqnarray}
The polarization is estimated as the summation of local electric dipole moments divided by the volume as follows;
\begin{eqnarray}
\textbf{P}& = & \frac{\lambda}{2NV} \sum_{i}^{2N}\sum_{j}^{4}  \left(  \textbf{S}_i  \cdot \textbf{e}_{ij}\right) ^2  \textbf{e}_{ij}\nonumber\\
&=&\frac{\lambda}{2V}\sum_{j=1}^{4} \left[ \left\lbrace  \left(  \textbf{S}_\mathrm{A}  \cdot \textbf{e}_{Aj}\right) ^2 + \left( \textbf{S}_\mathrm{B}  \cdot \textbf{e}_{Aj}\right) ^2 \right\rbrace \textbf{e}_{Aj} \right. \nonumber\\
&&\left. + \left\lbrace  \left(  \textbf{S}_\mathrm{A}  \cdot \textbf{e}_{Bj}\right) ^2 + \left( \textbf{S}_\mathrm{B}  \cdot \textbf{e}_{Bj}\right) ^2 \right\rbrace \textbf{e}_{Bj} \right] \nonumber\\
&=&\frac{8dfl\lambda}{V}
\begin{pmatrix}
S_{\mathrm{A}}^yS_{\mathrm{A}}^z + S_{\mathrm{B}}^yS_{\mathrm{B}}^z  \\
S_{\mathrm{A}}^xS_{\mathrm{A}}^z + S_{\mathrm{B}}^xS_{\mathrm{B}}^z  \\
S_{\mathrm{A}}^xS_{\mathrm{A}}^y + S_{\mathrm{B}}^xS_{\mathrm{B}}^y
\end{pmatrix}.
\label{eq:Ptot}
\end{eqnarray}
The effect of inter layer antiferromagnetic stacking is included in this formula. We introduce the ferromagnetic vector $\textbf{S}_{\mathrm{F}}$ and the antiferromagnetic vector $\textbf{S}_{\mathrm{AF}}$;
\begin{equation}
\textbf{S}_{\mathrm{F}} = 
\begin{pmatrix}
S_{\mathrm{A}}^x + S_{\mathrm{B}}^x \\
S_{\mathrm{A}}^y + S_{\mathrm{B}}^y  \\ 
S_{\mathrm{A}}^z + S_{\mathrm{B}}^z \\
\end{pmatrix},
\label{eq:m}
\end{equation}
\begin{equation}
\textbf{S}_{\mathrm{AF}} = 
\begin{pmatrix}
S_{\mathrm{A}}^x - S_{\mathrm{B}}^x  \\
S_{\mathrm{A}}^y - S_{\mathrm{B}}^y  \\ 
S_{\mathrm{A}}^z - S_{\mathrm{B}}^z \\
\end{pmatrix}.
\label{eq:l}
\end{equation}
With these vectors, the polarization can be expressed as
\begin{equation}
\textbf{P}^0 = \frac{4dfl\lambda}{V}
\begin{pmatrix}
S_{\mathrm{F}}^yS_{\mathrm{F}}^z + S_{\mathrm{AF}}^yS_{\mathrm{AF}}^z  \\
S_{\mathrm{F}}^xS_{\mathrm{F}}^z + S_{\mathrm{AF}}^xS_{\mathrm{AF}}^z  \\
S_{\mathrm{F}}^xS_{\mathrm{F}}^y + S_{\mathrm{AF}}^xS_{\mathrm{AF}}^y
\end{pmatrix}.
\label{eq:P^0}
\end{equation}

In order to compare with the experimentally observed polarization and estimate the coupling constant $\lambda$, we calculate the magnetic structure at finite temperature with use of molecular field approach. The magnitude of spin is expressed as the thermodynamical average $\left\langle \textbf{S}_\mathrm{A} \right\rangle$ and $\left\langle \textbf{S}_\mathrm{B} \right\rangle$.
\begin{eqnarray}
\left\langle \textbf{S}_i \right\rangle &=& \bar{S}\left( \cos\theta_i', \sin\theta_i', 0 \right)
\label{eq:barS}
\end{eqnarray}
Here $i = \mathrm{A},  \mathrm{B}$ and $\left| \left\langle \textbf{S}_\mathrm{A} \right\rangle \right| = \left| \left\langle \textbf{S}_\mathrm{B} \right\rangle \right| = \bar{S}$. From the mean-field approximation, the Hamiltonian is
\begin{eqnarray}
\mathcal{H} &=& \mathcal{H}_\mathrm{A} + \mathcal{H}_\mathrm{B},\\
\mathcal{H}_\mathrm{A} &=& \sum_{i}\left( 4J\left\langle \textbf{S}_\mathrm{B} \right\rangle + \mathrm{g\mu_\mathrm{B}} \mu_0 \textbf{H}  \right) \cdot \textbf{S}_{\mathrm{A}},\\
\mathcal{H}_\mathrm{B} &=& \sum_{i}\left( 4J\left\langle \textbf{S}_\mathrm{A} \right\rangle + g\mu_\mathrm{B} \mu_0 \textbf{H}  \right) \cdot \textbf{S}_{\mathrm{B}}.
\label{eq:Hamiltonian2}
\end{eqnarray}
The effective magnetic fields are
\begin{equation}
\textbf{H}_{\mathrm{eff},\mathrm{A}} = \frac{4J}{g\mu_\mathrm{B}}\left\langle \textbf{S}_\mathrm{B} \right\rangle + \mu_0 \textbf{H},
\label{eq:EwithT}
\end{equation}
\begin{equation}
\textbf{H}_{\mathrm{eff},\mathrm{B}} = \frac{4J}{g\mu_\mathrm{B}}\left\langle \textbf{S}_\mathrm{A} \right\rangle + \mu_0 \textbf{H}.
\label{eq:EwithT2}
\end{equation}
Because the magnetic torques are zero at steady state,
\begin{equation}
\textbf{H}_{\mathrm{eff},\mathrm{A}}\times g\mu_\mathrm{B} \left\langle \textbf{S}_\mathrm{A} \right\rangle = \textbf{H}_{\mathrm{eff},\mathrm{B}}\times g\mu_\mathrm{B}\left\langle \textbf{S}_\mathrm{B} \right\rangle = 0.
\label{eq:Magtorque}
\end{equation}
Thus the direction of spins is determined as follows;
\begin{equation}
2\theta' =  \theta_\mathrm{A}' - \theta_\mathrm{B}'\hspace{0.5cm}\left(\theta_\mathrm{A}' > \theta_\mathrm{B}' \right),
\label{eq:EwithT3}
\end{equation}
\begin{equation}
\cos\theta' = \frac{h}{8J\bar{S}},
\label{eq:theta'1}
\end{equation}
\begin{equation}
\theta_\mathrm{A}' = \theta_\mathrm{H} + \theta' + \pi,\hspace{0.5cm} \theta_\mathrm{B}' = \theta_\mathrm{H} - \theta' + \pi.
\label{eq:theta'2}
\end{equation}
The thermodynamical average of magnitude of spin $\bar{S}$ is expressed as follows;
\begin{eqnarray}
\bar{S} & = & S B_s\left[\frac{ \textbf{H}_{\mathrm{eff},\mathrm{A}}\cdot g\mu_\mathrm{B} \left\langle \textbf{S}_\mathrm{A} \right\rangle }{k_\mathrm{B}T} \right] \nonumber\\
&=& S B_s\left[ - \frac{4J\left\lbrace  2\left( h/8J \right) ^2 - \bar{S}^2 \right\rbrace   - h^2/8J}{k_BT}S \right].
\label{eq:thermoS}
\end{eqnarray}
Here $k_B$ is the Boltzmann constant and $B_s[x]$ is the Brillouin function,
\begin{eqnarray}
B_s\left[ x \right] &=& \frac{2S+1}{2S}\coth\left( \frac{2S+1}{2S}x \right) - \frac{1}{2S}\coth\left( \frac{x}{2S} \right).
\label{eq:Bs}
\end{eqnarray}

From Eq. (\ref{eq:thermoS}), we can numerically obtain the $h$ dependence of $\bar{S}$. The $h$ dependence of $\theta'$ is also obtained by Eq. (S20). In the magnetic field along [110] ($\theta_\mathrm{H} = \pi/4$),
\begin{eqnarray}
\left\langle  \textbf{S}_\mathrm{A} \right\rangle  =  \frac{\bar{S}}{\sqrt{2}} 
\begin{pmatrix}
-\cos\theta' + \sin\theta' \\
-\cos\theta' - \sin\theta'  \\
0
\end{pmatrix},\hspace{0.5cm}
\left\langle  \textbf{S}_\mathrm{B} \right\rangle  =  \frac{\bar{S}}{\sqrt{2}}
\begin{pmatrix}
-\cos\theta' -\sin\theta' \\
-\cos\theta' + \sin\theta'  \\
0
\end{pmatrix}, 
\label{eq:SB}
\end{eqnarray}
\begin{eqnarray}
\left\langle  \textbf{S}_\mathrm{F} \right\rangle  =  -\sqrt{2}\bar{S}\cos\theta' 
\begin{pmatrix}
1 \\
1 \\
0
\end{pmatrix},\hspace{0.5cm}  
\left\langle  \textbf{S}_\mathrm{AF} \right\rangle  =  \sqrt{2}\bar{S}\sin\theta' 
\begin{pmatrix}
1 \\
-1 \\
0
\end{pmatrix}.
\label{eq:SFSAF}
\end{eqnarray}
Thus the polarization is
\begin{equation}
\textbf{P} = \frac{8dfl\lambda}{V}\bar{S}^2
\begin{pmatrix}
0 \\
0 \\
2\cos^2\theta' - 1 
\end{pmatrix}.
\label{eq:P0}
\end{equation}
Figure S2 compares the obtained polarization and experimental data[3].  Here we used parameters, $S=5/2$, $T = 1.8$ K, $V = 8.5022\times8.5022\times5.5244\times10^{-30}$ m$^3$, and $4JS/g\mu_\mathrm{B} = \mu_0H_\mathrm{E} =$ 4.67 T. The $h$-dependences are similar to each other. From the comparison, we obtained $|\lambda|$ is estimated as $9\times10^{-35}$Cm.

\begin{figure}[tbh]
	\centering
	\includegraphics[width=0.4\linewidth]{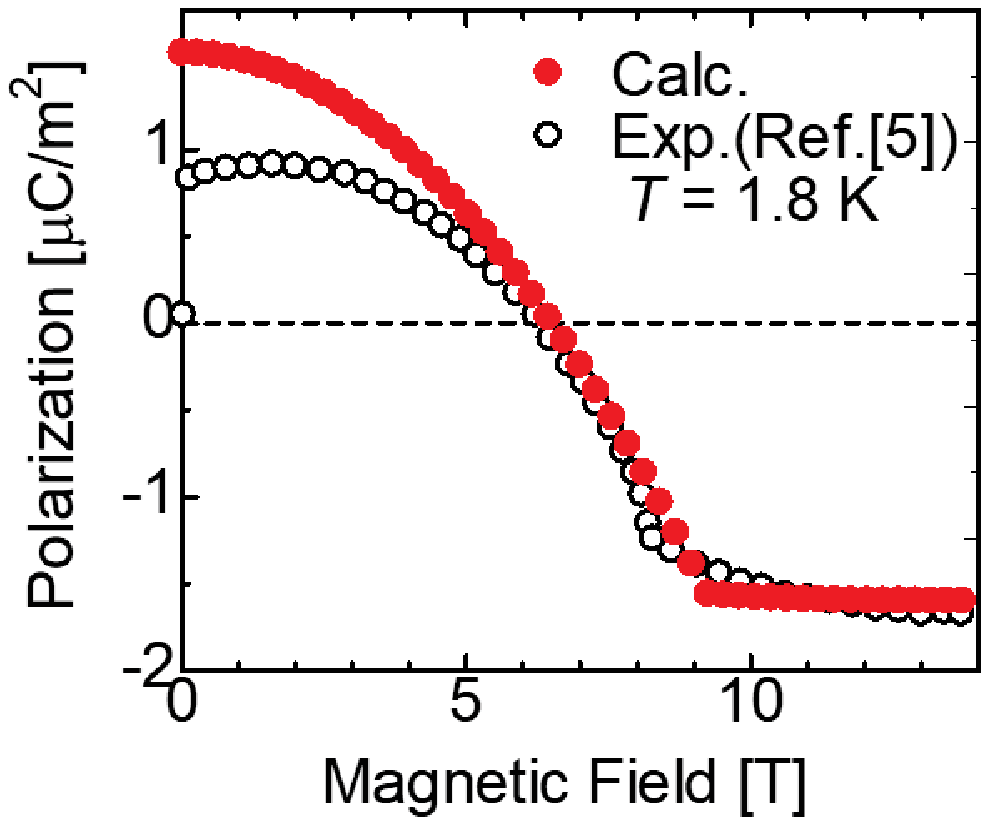}
	\caption{
		Polarization experimentally obtained by Murakawa {\it et al.}\cite{Murakawa2012} and calculated polarization based on Eq. (\ref{eq:P0}) with $|\lambda|$ = $9\times10^{-35}$ Cm.
	}
	\label{fig:P-H}
\end{figure}

\section*{SIII. Antiferromagnetic magnon modes}

In this section, we discuss the antiferromagnetic magnon modes. Finite temperature effect is neglected for simplicity. First, we introduced the coordinate system along the spin direction. The spin coordinate system is rotated so that the $x$-axis is aligned with the direction of ordered spin moments by the unitary operator
\begin{equation}
U = \exp \left( -i\sum_{i}\theta_i S_i^z \right).
\label{eq:U1}
\end{equation}
The spin moments in the rotated system ($\tilde{\textbf{S}}_i$) are 
\begin{equation}
U^\dagger\textbf{S}_i U \equiv \tilde{\textbf{S}}_i = R_z\left( \theta_i\right) \textbf{S}_i,
\label{eq:USU}
\end{equation}
where
\begin{equation}
R_z\left( \theta_i\right) = 
\begin{pmatrix}
\cos\theta_i & -\sin\theta_i & 0 \\
\sin\theta_i & \cos\theta_i & 0 \\
0 & 0 & 1
\end{pmatrix}.
\label{eq:Rz}
\end{equation}
The Hamiltonian (Eq. (\ref{eq:Hamiltonian})) is transformed by $U$ into
\begin{equation}
\begin{split}
\tilde{\mathcal{H}} &= J \sum_{<i,j>}\left\lbrace \left(\tilde{S}_i^x\tilde{S}_j^x + \tilde{S}_i^y\tilde{S}_j^y \right) \cos\left( \theta_i - \theta_j \right) + \tilde{S}_i^z\tilde{S}_j^z + \left[\tilde{\textbf{S}}_i \times \tilde{\textbf{S}}_j \right] ^z \sin\left(\theta_i - \theta_j \right)  \right\rbrace
\\
&\hspace{0.5cm} + K\sum_{i}\left(\tilde{S}_i^z \right) ^2 +h\sum_{i}\left\lbrace \tilde{S}_i^x \cos\left( \theta_i - \theta_\mathrm{H}\right)  - \tilde{S}_i^y \sin\left(\theta_i - \theta_\mathrm{H} \right)  \right\rbrace.
\end{split}
\label{eq:UHU}
\end{equation}
In the rotated system, the Holstein-Primakoff (H-P) transformations are
\begin{equation}
\tilde{S}_i^x = S -a_i^\dagger a_i,\hspace{0.3cm} \tilde{S}_i^y = \sqrt{\frac{S}{2}}\left(a_i + a_i^\dagger \right) + O\left(S^{-\frac{1}{2}} \right),\hspace{0.3cm} \tilde{S}_i^z = -i\sqrt{\frac{S}{2}}\left( a_i - a_i^\dagger \right) + O\left( S^{-\frac{1}{2}} \right),    
\label{eq:HP1}
\end{equation}
\begin{equation}
\tilde{S}_j^x = S -b_j^\dagger b_j,\hspace{0.3cm} \tilde{S}_j^y = \sqrt{\frac{S}{2}}\left(b_j + b_j^\dagger \right) + O\left(S^{-\frac{1}{2}} \right),\hspace{0.3cm} \tilde{S}_j^z = -i\sqrt{\frac{S}{2}}\left( b_j - b_j^\dagger \right) + O\left( S^{-\frac{1}{2}} \right).    
\label{eq:HP2}
\end{equation}
Here $a_i$, $b_j$ and $a_i^\dagger$, $b_j^\dagger$ are the boson annihilation and creation operators, respectively. In this supplemental information, we discuss the magnon modes coupled to the microwave. The microwave wavelength is fairly long compared with the atomic distance. The coupled magnon modes can be regarded as spatially uniform. Therefore, we assume that $a_i$,$a_i^\dagger$, $b_i$, and $b_i^\dagger$ are independent of atomic site indicated by suffix $i$. Hereafter, we omit the suffix. Then the H-P transformed Hamiltonian becomes 
\begin{equation}
\tilde{\mathcal{H}} = \bar{E} + \frac{1}{2}\Psi^\dagger \mathcal{H}_M \Psi + O\left(S^\frac{1}{2} \right).
\label{eq:UHU2}
\end{equation}
Here,
\begin{equation}
\bar{E} = 4JNS\left(S+1\right) \cos 2\theta - hN\left( 2S + 1 \right) \cos\theta,
\label{eq:Ebar}
\end{equation}
\begin{equation}
\Psi^\dagger = \left( a^\dagger,\hspace{0.2cm} b^\dagger,\hspace{0.2cm} a,\hspace{0.2cm}b \right),
\label{eq:Psi}
\end{equation}
\begin{equation}
\mathcal{H}_M = 
\begin{pmatrix}
4JS+KS & 4JS\cos^2\theta & -KS & -4JS\sin^2\theta \\
4JS\cos^2\theta & 4JS+KS & -4JS\sin^2\theta & -KS \\
-KS & -4JS\sin^2\theta & 4JS+KS & 4JS\cos^2\theta \\
-4JS\sin^2\theta & -KS & 4JS\cos^2\theta & 4JS+KS
\end{pmatrix}.
\label{eq:H_M}
\end{equation}
The magnon energy $\omega_n$ is obtained by the secular equation 
\begin{equation}
\Sigma^z \mathcal{H}_M \textbf{u}_n = \omega_n \textbf{u}_n,
\label{eq:ee1}
\end{equation}
where
\begin{equation}
\Sigma^z = 
\begin{pmatrix}
1 & 0 & 0 & 0\\
0 & 1 & 0 & 0\\
0 & 0 & -1 & 0\\
0 & 0 & 0 & -1
\end{pmatrix}.
\label{eq:Sigmaz}
\end{equation}
The eigenvalues are obtained as:
\begin{equation}
\omega_{1} = 8JS\cos\theta\sqrt{1+\frac{K}{4J}} = g\mu_\mathrm{B}\mu_0H\sqrt{1+\frac{H_\mathrm{A}}{2H_\mathrm{E}}},
\label{eq:Omega-}
\end{equation}
\begin{equation}
\omega_{2} = \sqrt{16JSKS\left( 1-\cos^2\theta\right)  } = g\mu_\mathrm{B}\mu_0 \sqrt{2H_\mathrm{E}H_\mathrm{A}\left(1-\left(\frac{H}{2H_\mathrm{E}} \right) ^2 \right) }.
\label{eq:Omega+}
\end{equation}
Here the exchange field $H_\mathrm{E}$ and the magnetic anisotropy field $H_\mathrm{A}$ are defined as
\begin{equation}
H_\mathrm{E} = \frac{4JS}{g\mu_\mathrm{B}\mu_0},\hspace{0.5cm} H_\mathrm{A} =\frac{2KS}{g\mu_\mathrm{B}\mu_0}.
\label{eq:HAHE}
\end{equation}
The diagonalized Hamiltonian is obtained by the Bogoliubov transformation
\begin{equation}
\begin{pmatrix}
a\\
b \\
a^\dagger \\
b^\dagger
\end{pmatrix}
= \frac{1}{\sqrt{2}}
\begin{pmatrix}
\cosh\phi_{2} & \cosh\phi_{1} & -\sinh\phi_{2} & \sinh\phi_{1} \\
-\cosh\phi_{2} & \cosh\phi_{1} & \sinh\phi_{2} & \sinh\phi_{1} \\
-\sinh\phi_{2} & \sinh\phi_{1} & \cosh\phi_{2} & \cosh\phi_{1} \\
\sinh\phi_{2} & \sinh\phi_{1} & -\cosh\phi_{2} & \cosh\phi_{1}
\end{pmatrix}
\begin{pmatrix}
\alpha \\
\beta \\
\alpha^\dagger \\
\beta^\dagger
\end{pmatrix},
\label{eq:BT}
\end{equation}
where
\begin{equation}
\cosh\phi_{2} = \sqrt{\frac{4JS\left( 1-\cos^2\theta \right) + KS }{2\omega_{2}} + \frac{1}{2}},
\label{eq:coshphi+}
\end{equation}
\begin{equation}
\sinh\phi_{2} =  \sqrt{\frac{4JS\left( 1-\cos^2\theta \right) + KS }{2\omega_{2}} -\frac{1}{2}}, \hspace{0.5cm} \left( h \leq 8JS\sqrt{1-\frac{K}{4J}} \right), 
\label{eq:sinhphi+}
\end{equation}
\begin{equation}
\cosh\phi_{1} = \sqrt{\frac{4JS\left( 1+\cos^2\theta \right) + KS }{2\omega_{1}} + \frac{1}{2}},
\label{eq:coshphi-}
\end{equation}
\begin{equation}
\sinh\phi_{1} = \sqrt{\frac{4JS\left( 1+\cos^2\theta \right) + KS }{2\omega_{1}} - \frac{1}{2}}.
\label{eq:sinhphi-}
\end{equation}
\begin{equation}
\tilde{\mathcal{H}} = \bar{E} + \omega_{2}\left(\alpha^\dagger\alpha + \frac{1}{2} \right) + \omega_{1} \left( \beta^\dagger \beta + \frac{1}{2} \right) + O\left( S^\frac{1}{2}\right),   
\label{eq:UHU3}
\end{equation}
With use of creation and annihilation operators, we can expressed $\textbf{S}_{\mathrm{A}}$ and $\textbf{S}_{\mathrm{B}}$ as
\begin{equation}
\textbf{S}_{\mathrm{A}} = \textbf{m}_\mathrm{A}S -\textbf{m}_\mathrm{A} a^\dagger a + \sqrt{\frac{S}{2}}\left\lbrace\frac{\partial \textbf{m}_\mathrm{A}}{\partial\theta_\mathrm{A}}\left( a + a^\dagger \right) - i \hat{\textbf{z}}\left( a - a^\dagger \right)   \right\rbrace + O\left( S^{-\frac{1}{2}} \right) ,
\label{eq:USiU}
\end{equation}
\begin{equation}
\textbf{S}_{\mathrm{B}} = \textbf{m}_\mathrm{B}S -\textbf{m}_\mathrm{B} b^\dagger b + \sqrt{\frac{S}{2}}\left\lbrace\frac{\partial \textbf{m}_\mathrm{B}}{\partial\theta_\mathrm{B}}\left( b + b^\dagger \right) - i \hat{\textbf{z}}\left( b - b^\dagger \right)   \right\rbrace + O\left( S^{-\frac{1}{2}} \right) ,
\label{eq:USjU}
\end{equation}
\begin{equation}
\textbf{m}_\mathrm{A} = \left( \cos\theta_\mathrm{A}, \sin\theta_\mathrm{A}, 0 \right),\hspace{0.5cm} \textbf{m}_\mathrm{B} = \left( \cos\theta_\mathrm{B}, \sin\theta_\mathrm{B}, 0 \right) . 
\label{eq:m1m2}
\end{equation}
In the case of $\theta_\mathrm{H} = 0$,
\begin{equation}
\frac{\partial \textbf{m}_\mathrm{A}}{\partial\theta_\mathrm{A}} =
\begin{pmatrix}
\sin\theta\\
-\cos\theta\\
0
\end{pmatrix}
,\hspace{0.5cm} \frac{\partial\textbf{m}_\mathrm{B}}{\partial\theta_\mathrm{B}} =
\begin{pmatrix}
-\sin\theta\\
-\cos\theta\\
0
\end{pmatrix},
\label{eq:m1m2_2}
\end{equation}
\begin{eqnarray}
\textbf{S}_{\mathrm{F}} &=&  \textbf{S}_{\mathrm{A}} + \textbf{S}_{\mathrm{B}} \nonumber
\\
&=& \sqrt{\frac{S}{N}} 
\begin{pmatrix}
\sin\theta\left( \cosh\phi_{2} - \sinh\phi_{2} \right) \left( \alpha + \alpha^\dagger \right)  \\
-\cos\theta \left( \cosh\phi_{1} + \sinh\phi_{1} \right) \left( \beta + \beta^\dagger \right)  \\
-i \left( \cosh\phi_{1} - \sinh\phi_{1} \right) \left( \beta - \beta^\dagger \right)  
\end{pmatrix}
\nonumber\\
&& - 2S
\begin{pmatrix}
\cos\theta\\
0  \\
0
\end{pmatrix}\nonumber\\
&& + \left(\mathrm{2nd\hspace{0.1cm}order\hspace{0.1cm}terms\hspace{0.1cm}of}\hspace{0.1cm}\alpha, \beta \right) +O\left(S^{-\frac{1}{2}} \right).  
\label{eq:UStotU}
\end{eqnarray}
The dynamical and static components of $\textbf{S}_{\mathrm{F}}$ ($\textbf{S}_{F}^\omega$ and $\textbf{S}_{F}^0$) are, respectively, expressed by the first and second terms as follows:
\begin{eqnarray}
\textbf{S}_{F}^\omega&=& \sqrt{\frac{S}{N}} 
\begin{pmatrix}
\sin\theta\left( \cosh\phi_{2} - \sinh\phi_{2} \right) \left( \alpha + \alpha^\dagger \right)  \\
-\cos\theta \left( \cosh\phi_{1} + \sinh\phi_{1} \right) \left( \beta + \beta^\dagger \right)  \\
-i \left( \cosh\phi_{1} - \sinh\phi_{1} \right) \left( \beta - \beta^\dagger \right)  
\end{pmatrix}, \\
\textbf{S}_{F}^0&=& - 2S
\begin{pmatrix}
\cos\theta\\
0  \\
0
\end{pmatrix}.
\label{eq:SFw-0}
\end{eqnarray}
Similarly,
\begin{eqnarray}
\textbf{S}_{\mathrm{AF}} &=& \textbf{S}_{\mathrm{A}} - \textbf{S}_{\mathrm{B}} \nonumber
\\
&=& \sqrt{\frac{S}{N}} 
\begin{pmatrix}
\sin\theta\left( \cosh\phi_{1} + \sinh\phi_{1} \right) \left( \beta + \beta^\dagger \right)  \\
-\cos\theta \left( \cosh\phi_{2} - \sinh\phi_{2} \right) \left( \alpha + \alpha^\dagger \right)  \\
-i \left( \cosh\phi_{2} + \sinh\phi_{2} \right) \left( \alpha - \alpha^\dagger \right)  
\end{pmatrix}
\nonumber\\
&& - 2S
\begin{pmatrix}
0\\
\sin\theta  \\
0
\end{pmatrix}\nonumber\\
&& + \left(\mathrm{2nd\hspace{0.1cm}order\hspace{0.1cm}terms\hspace{0.1cm}of}\hspace{0.1cm}\alpha, \beta \right) +O\left(S^{-\frac{1}{2}} \right).
\label{eq:USAFU}
\end{eqnarray}
The dynamical and static components of $\textbf{S}_{\mathrm{AF}}$ ($\textbf{S}_{AF}^\omega$ and $\textbf{S}_{AF}^0$) are, respectively, defined by the first and second terms as follows:
\begin{eqnarray}
\textbf{S}_{AF}^\omega&=& \sqrt{\frac{S}{N}} 
\begin{pmatrix}
\sin\theta\left( \cosh\phi_{1} + \sinh\phi_{1} \right) \left( \beta + \beta^\dagger \right)  \\
-\cos\theta \left( \cosh\phi_{2} - \sinh\phi_{2} \right) \left( \alpha + \alpha^\dagger \right)  \\
-i \left( \cosh\phi_{2} + \sinh\phi_{2} \right) \left( \alpha - \alpha^\dagger \right)  
\end{pmatrix},\\
\textbf{S}_{AF}^0&=& - 2S
\begin{pmatrix}
0\\
\sin\theta  \\
0
\end{pmatrix}.
\label{eq:SAFw-0}
\end{eqnarray}

In the case of $\theta_\mathrm{H} = 3\pi/4$,
\begin{equation}
\frac{\partial \textbf{m}_\mathrm{A}}{\partial\theta_\mathrm{A}} = \frac{1}{\sqrt{2}}
\begin{pmatrix}
\cos\theta - \sin\theta\\
\cos\theta + \sin\theta\\
0
\end{pmatrix}
,\hspace{0.5cm} \frac{\partial\textbf{m}_\mathrm{B}}{\partial\theta_\mathrm{B}} = \frac{1}{\sqrt{2}}
\begin{pmatrix}
\cos\theta + \sin\theta\\
\cos\theta - \sin\theta\\
0
\end{pmatrix},
\label{eq:m1m2_22}
\end{equation}
\begin{eqnarray}
\textbf{S}_{\mathrm{F}}  &=& \sqrt{\frac{S}{2N}} \times\nonumber\\
&&\hspace{-2cm}\begin{pmatrix}
-\sin\theta\left( \cosh\phi_{2} - \sinh\phi_{2} \right) \left( \alpha + \alpha^\dagger \right) +\cos\theta \left( \cosh\phi_{1} + \sinh\phi_{1} \right) \left( \beta + \beta^\dagger \right) \\
\sin\theta\left( \cosh\phi_{2} - \sinh\phi_{2} \right) \left( \alpha + \alpha^\dagger \right) + \cos\theta \left( \cosh\phi_{1} + \sinh\phi_{1} \right) \left( \beta + \beta^\dagger \right)  \\
-\sqrt{2}i \left( \cosh\phi_{1} - \sinh\phi_{1} \right) \left( \beta - \beta^\dagger \right)  
\end{pmatrix}
\nonumber\\
&& + \sqrt{2}S\cos\theta
\begin{pmatrix}
1\\
-1\\
0
\end{pmatrix}\nonumber\\
&& + \left(\mathrm{2nd\hspace{0.1cm}order\hspace{0.1cm}terms\hspace{0.1cm}of}\hspace{0.1cm}\alpha, \beta \right) +O\left(S^{-\frac{1}{2}} \right),  
\label{eq:UStotU2}
\end{eqnarray}
\begin{eqnarray}
\textbf{S}_{\mathrm{AF}}  &=& \sqrt{\frac{S}{2N}} \times\nonumber\\
&&\hspace{-2cm}\begin{pmatrix}
\cos\theta \left( \cosh\phi_{2} - \sinh\phi_{2} \right) \left( \alpha + \alpha^\dagger \right) -\sin\theta\left( \cosh\phi_{1} + \sinh\phi_{1} \right) \left( \beta + \beta^\dagger \right) \\
\cos\theta \left( \cosh\phi_{2} - \sinh\phi_{2} \right) \left( \alpha + \alpha^\dagger \right) + \sin\theta \left( \cosh\phi_{1} + \sinh\phi_{1} \right) \left( \beta + \beta^\dagger \right) \\
-\sqrt{2}i \left( \cosh\phi_{2} + \sinh\phi_{2} \right) \left( \alpha - \alpha^\dagger \right)  
\end{pmatrix}
\nonumber\\
&& + \sqrt{2}S\sin\theta
\begin{pmatrix}
1\\
1\\
0
\end{pmatrix}.
\label{eq:USAFU2}
\end{eqnarray}

\section*{SIV. Microwave non-reciprocity}

\subsection{Dynamical Susceptibility tensors}
In this section, we discuss dynamical susceptibility tensors for the estimation of microwave non-reciprocity in the later section. For the magnetoelectric substance, the oscillating electric and magnetic flux densities ($\textbf{D}^\omega=(D^\omega_x, D^\omega_y, D^\omega_z), \textbf{B}^\omega=(B^\omega_x,B^\omega_y, B^\omega_z)$) in oscillating electric and magnetic fields ($\textbf{E}^\omega=(E^\omega_x, E^\omega_y, E^\omega_z)$, $\textbf{H}^\omega=(H^\omega_x,H^\omega_y, H^\omega_z)$) can be expressed as:
\begin{equation}
D^\omega_i=\varepsilon_0(\varepsilon_\infty+\chi^{ee}_{ij})E_j^\omega+\sqrt{\varepsilon_0\mu_0}\chi^{em}_{ij}H_j,
\label{eq:Domega}
\end{equation}
\begin{equation}
B^\omega_i=\mu_0(1+\chi^{mm}_{ij})H_j^\omega+\sqrt{\varepsilon_0\mu_0}\chi^{me}_{ij}E_j,
\label{eq:Bomega}
\end{equation}
where $\chi^{mm}, \chi^{ee}$, $\chi^{em}$ and $\chi^{me}$ are magnetic, electric, electromagnetic, and magnetoelectric dynamical tensors, respectively. $\varepsilon_0$ is the permittivity in vacuum. $\varepsilon_\infty$ is the relative permittivity at high frequency. According to ref. 9\cite{Su2012}, $\varepsilon_\infty \approx 14$. The nonzero component of  these dynamical susceptibility tensors can be determined by the symmetry analysis\cite{Birss,Graham}. Let us discuss them under $\textbf{H}\parallel$[100] and $\textbf{H}\parallel$[$1\bar{1}0$] corresponding to the experiments. The magnetic point groups are $22'2'$ and $m'm2'$ for $\textbf{H}\parallel$[100] and $\textbf{H}\parallel$[$1\bar{1}0$], respectively. Therefore, for $\textbf{H}\parallel$[100],
\begin{equation}
\chi^{mm} = 
\begin{pmatrix}
\chi_{xx}^{mm} & 0 & 0 \\
0 & \chi_{yy}^{mm} & \chi_{yz}^{mm} \\
0 & -\chi_{yz}^{mm} & \chi_{zz}^{mm}
\end{pmatrix}
,\hspace{0.5cm}
\chi^{ee} =
\begin{pmatrix}
\chi_{xx}^{ee} & 0 & 0 \\
0 & \chi_{yy}^{ee} & \chi_{yz}^{ee} \\
0 & -\chi_{yz}^{ee} & \chi_{zz}^{ee}
\end{pmatrix},\nonumber
\label{eq:chieemm}
\end{equation}
\begin{equation}
\chi^{me} = 
\begin{pmatrix}
\chi_{xx}^{me} & 0 & 0 \\
0 & \chi_{yy}^{me} & \chi_{yz}^{me} \\
0 & \chi_{zy}^{me} & \chi_{zz}^{me}
\end{pmatrix}
,\hspace{0.5cm}
\chi^{em} =
\begin{pmatrix}
-\chi_{xx}^{me} & 0 & 0 \\
0 & -\chi_{yy}^{me} & \chi_{zy}^{me} \\
0 & \chi_{yz}^{me} & -\chi_{zz}^{me}
\end{pmatrix},
\label{eq:chiemme}
\end{equation}
where $\textbf{x}\parallel [100], \textbf{y}\parallel [010]$ and $\textbf{z}\parallel [001]$. For $\textbf{H}\parallel$[$1\bar{1}0$],
\begin{equation}
\chi^{mm} = 
\begin{pmatrix}
\chi_{xx}^{mm} & 0 & \chi_{xz}^{mm} \\
0 & \chi_{yy}^{mm} & 0 \\
-\chi_{xz}^{mm} & 0 & \chi_{zz}^{mm}
\end{pmatrix}
,\hspace{0.5cm}
\chi^{ee} =
\begin{pmatrix}
\chi_{xx}^{ee} & 0 & \chi_{xz}^{ee} \\
0 & \chi_{yy}^{ee} & 0 \\
-\chi_{xz}^{ee} & 0 & \chi_{zz}^{ee}
\end{pmatrix},\nonumber
\label{eq:chieemm2}
\end{equation}
\begin{equation}
\chi^{me} = 
\begin{pmatrix}
0 & 0 & \chi_{xz}^{me} \\
0 & 0 & \chi_{yz}^{me} \\
\chi_{zx}^{me} & \chi_{zy}^{me} & 0
\end{pmatrix}
,\hspace{0.5cm}
\chi^{em} =
\begin{pmatrix}
0 & 0 & -\chi_{zx}^{me} \\
0 & 0 & \chi_{zy}^{me} \\
-\chi_{xz}^{me} & \chi_{yz}^{me} & 0
\end{pmatrix},
\label{eq:chiemme2}
\end{equation}
where $\textbf{x}\parallel [110], \textbf{y}\parallel [001]$ and $\textbf{z}\parallel [1\bar{1}0]$.

The dynamical susceptibility tensors at $T = 0$ are obtained by the Kubo formula as follows;
\begin{equation}
\chi_{\beta\gamma}^{me} = \frac{NV}{\hbar}\sqrt{\frac{\mu_0}{\varepsilon_0}}\sum_n\frac{\left\langle 0 \left| \Delta M_\beta \right| n  \right\rangle \left\langle n \left| \Delta P_\gamma \right| 0  \right\rangle }{\omega - \omega_n + i\delta},
\label{eq:chime}
\end{equation}
\begin{equation}
\chi_{\beta\gamma}^{em} = \frac{NV}{\hbar}\sqrt{\frac{\mu_0}{\varepsilon_0}}\sum_n\frac{\left\langle 0 \left| \Delta P_\beta \right| n  \right\rangle \left\langle n \left| \Delta M_\gamma \right| 0  \right\rangle }{\omega - \omega_n + i\delta},
\label{eq:chiem}
\end{equation}
\begin{equation}
\chi_{\beta\gamma}^{mm} = \frac{NV}{\hbar}\mu_0\sum_n\frac{\left\langle 0 \left| \Delta M_\beta \right| n  \right\rangle \left\langle n \left| \Delta M_\gamma \right| 0  \right\rangle }{\omega - \omega_n + i\delta},
\label{eq:chimm}
\end{equation}
\begin{equation}
\chi_{\beta\gamma}^{ee} = \frac{NV}{\hbar}\frac{1}{\varepsilon_0}\sum_n\frac{\left\langle 0 \left| \Delta P_\beta \right| n  \right\rangle \left\langle n \left| \Delta P_\gamma \right| 0  \right\rangle }{\omega - \omega_n + i\delta},
\label{eq:chiee}
\end{equation}
where $\left| 0  \right\rangle$ is the ground state and $\left| n  \right\rangle$ is the magnon excited state. Here, $\Delta \textbf{M}$ and $\Delta \textbf{P}$ are, respectively, the dynamical polarization and magnetization induced by the magnons expressed as follows;
\begin{equation}
\Delta\textbf{M} = -\frac{1}{2NV}\sum_{i}^{2N}g\mu_\mathrm{B}\Delta \textbf{S}_i \simeq -\frac{1}{V}g\mu_\mathrm{B} \textbf{S}_{\mathrm{F}}^\omega,
\label{eq:DeltaM}
\end{equation}
\begin{eqnarray}
\Delta \textbf{P} &=& \frac{\lambda}{NV}\sum_{i}^{2N}\sum_{j}^{4}\left(\textbf{S}_i\cdot \textbf{e}_{ij} \right) \left( \Delta\textbf{S}_i\cdot \textbf{e}_{ij} \right) \textbf{e}_{ij}\nonumber\\
&=& \frac{4dfl\lambda}{V} 
\begin{pmatrix}
S_{\mathrm{F},y}^{0}S_{\mathrm{F},z}^{\omega} + S_{\mathrm{F},y}^{\omega}S_{\mathrm{F},z}^{0} + S_{\mathrm{AF},y}^{0}S_{\mathrm{AF},z}^{\omega} + S_{\mathrm{AF},y}^{\omega}S_{\mathrm{AF},z}^{0}  \\
S_{\mathrm{F},x}^{0}S_{\mathrm{F},z'}^{\omega} + S_{\mathrm{F},x}^{\omega}S_{\mathrm{F},z'}^{0} + S_{\mathrm{AF},x}^{0}S_{\mathrm{AF},z'}^{\omega} + S_{\mathrm{AF},x}^{\omega}S_{\mathrm{AF},z}^{0}  \\
S_{\mathrm{F},x}^{0}S_{\mathrm{F},y}^{\omega} + S_{\mathrm{F},x}^{\omega}S_{\mathrm{F},y}^{0} + S_{\mathrm{AF},x}^{0}S_{\mathrm{AF},y}^{\omega} + S_{\mathrm{AF},x}^{\omega}S_{\mathrm{AF},y}^{0}
\end{pmatrix}.
\label{eq:DeltaP2}
\end{eqnarray}

\subsection{Microwave non-reciprocity in coplanar waveguide}
In order to theoretically obtain the microwave non-reciprocity, we should estimate the damping rate of microwave in the microwave coplanar waveguide with sample. We assume that $x'y'z'$-coordinate is fixed to the microwave wave guide. The $x'$-direction is along the microwave propagation direction. $y'$ is parallel to the coplanar pattern but perpendicular to $x'$. The $z'$ direction is perpendicular to the coplanar pattern. In our experimental setup, the microwave is composed of two linearly polarized waves (polarization 1: $\textbf{E}^\omega \parallel \textbf{z}'$, $\textbf{H}^\omega \parallel \textbf{y}'$) and (polarization 2: $\textbf{E}^\omega \parallel \textbf{y}'$, $\textbf{H}^\omega \parallel \textbf{z}'$). For simplicity, we assume the two polarizations are equally mixed. We also assume that the linear polarization is approximately maintained in the substance. In order to estimate the refractive index for the polarization 1 ($\textbf{E}^\omega \parallel \textbf{z}', \textbf{H}^\omega \parallel \textbf{y}'$), We put $E^\omega_{x'} = E^\omega_{y'} = H^\omega_{x'} = H^\omega_{z'} = 0$, $E^\omega_{z'} = \left| E^\omega_{z'} \right| \exp\left[ i(kx'-\omega t)\right]$, $H^\omega_{y'} = \left| H^\omega_{y'} \right| \exp \left[ i(kx'-\omega t)\right]$ into the Maxwell equations, and obtain 
\begin{equation}
-kE_{z'}^\omega = \omega \left\lbrace  \left( 1 + \chi_{y'y'}^{mm}  \right)\mu_0H_{y'}^\omega +\chi_{y'z'}^{me}\sqrt{\varepsilon_0\mu_0}E_{z'}^\omega  \right\rbrace, 
\label{eq:kE}
\end{equation}
\begin{equation}
kH_{y'}^\omega = -\omega \left\lbrace \left( \varepsilon_\infty + \chi_{z'z'}^{ee} \right) \varepsilon_0 E_{z'}^\omega + \chi_{z'y'}^{em}\sqrt{\varepsilon_0\mu_0}H_{y'}^\omega  \right\rbrace.
\label{eq:kH}
\end{equation}
From the requirement of existence of solution other than $E^\omega_z = H^\omega_y = 0$, we get 
\begin{equation}
k = \omega \sqrt{\varepsilon_0\mu_0}\left( -\frac{\chi_{y'z'}^{me} + \chi_{z'y'}^{em}}{2} \pm \sqrt{ \left( \varepsilon_\infty + \chi_{z'z'}^{ee} \right)  \left( 1 + \chi_{y'y'}^{mm} \right)  } \right).
\label{eq:k}
\end{equation}
The magnitude of second term is much larger than that of first term. Therefore, the upper sign is corresponding to the $k>0$ solution while the lower sign to the $k<0$ solution. The difference of refractive indices $n$ for positive and negative $k$ is
\begin{equation}
\Delta n = - \left( \chi_{y'z'}^{me} + \chi_{z'y'}^{em} \right) .
\label{eq:Dn}
\end{equation}
The average of refractive indices is
\begin{equation}
\bar{n} = \sqrt{ \left( \varepsilon_\infty + \chi_{z'z'}^{ee} \right)  \left( 1 + \chi_{y'y'}^{mm} \right)  }.
\label{eq:nbar}
\end{equation}
Because the absorption coefficient $\alpha$ is expressed as $\omega\mathrm{Im}\left[ n \right]/c$, the difference of absorption coefficient is
\begin{equation}
\Delta \alpha_1 = -\frac{\omega}{c}\mathrm{Im}\left[ \chi_{y'z'}^{me} + \chi_{z'y'}^{em} \right],
\label{eq:Dalpha}
\end{equation}
and the average of absorption coefficient is
\begin{equation}
\bar{\alpha}_1 = \frac{\omega}{c}\mathrm{Im}\left[ \sqrt{ \left( \varepsilon_\infty + \chi_{z'z'}^{ee} \right)  \left( 1 + \chi_{y'y'}^{mm} \right)  } \right]. 
\label{eq:baralpha}
\end{equation}
The suffix "1" stands for the first polarization ($\textbf{E}^\omega \parallel \textbf{z}$, $\textbf{H}^\omega \parallel \textbf{y}$). 

On the other hand, for the polarization $\textbf{E}^\omega \parallel \textbf{y}', \textbf{H}^\omega \parallel \textbf{z}'$ (polarization 2), the microwave non-reciprocity and the average of microwave absorption are, respectively,
\begin{eqnarray}
\Delta\alpha_2 &=& \frac{\omega}{c}\mathrm{Im}\left[ \chi_{z'y'}^{me} + \chi_{y'z'}^{em} \right],\\
\bar{\alpha}_2 &=& \frac{\omega}{c}\mathrm{Im}\left[ \sqrt{ \left( \varepsilon_\infty + \chi_{y'y'}^{ee} \right)  \left( 1 + \chi_{z'z'}^{mm} \right)  } \right].
\label{eq:Dalpha2}
\end{eqnarray}
We assume the relative magnitude of the microwave non-reciprocity in our experiment is corresponding to
\begin{equation}
\frac{\Delta\alpha}{\bar{\alpha}} \simeq \frac{\Delta\alpha_1 + \Delta\alpha_2}{\bar{\alpha}_1 + \bar{\alpha}_2}.
\label{eq:Dabara}
\end{equation}
The microwave absorption spectrum is obtained from the absorption coefficients as follows;
\begin{eqnarray}
\Delta S_{12} + \Delta S_{21}& = &  -2\bar{\alpha}L \times 20\log_{10}e,\\
\Delta S_{12} - \Delta S_{21}& = &  -\Delta\alpha L \times 20\log_{10}e.
\label{eq:S-parameter_alpha}
\end{eqnarray}
Here $L$ is the propagation length of microwave in a sample. Thus the relative magnitude of the microwave non-reciprocity is equivalent to the experimental value.
\begin{equation}
2\frac{\Delta S_{12} - \Delta S_{21}}{\Delta S_{12} + \Delta S_{21}} = \frac{\Delta\alpha}{\bar{\alpha}}
\label{eq:exptheoDalp}
\end{equation}

\subsection{Microwave non-reciprocity for $\textbf{H}\parallel[100]$ and $\textbf{H}^\omega \perp [100]$}
In this subsection, we theoretically estimate the non-reciprocity for $\textbf{H}\parallel[100]$, $\textbf{H}^\omega \perp [100]$. The real and imaginary part of the dynamical susceptibilities are expressed as follows:
\begin{eqnarray}
\mathrm{Im}\left[ \chi_{y'z'}^{me} + \chi_{z'y'}^{em} \right]  &=& \frac{NV}{\hbar}\sqrt{\frac{\mu_0}{\varepsilon_0}}\sum_n\frac{-2\delta \left\langle 0 \left| \Delta M_{y'} \right| n  \right\rangle \left\langle n \left| \Delta P_z' \right| 0  \right\rangle}{\left( \omega - \omega_n \right) ^2 + \delta^2},\\
\mathrm{Im}\left[ \chi_{z'y'}^{me} + \chi_{y'z'}^{em} \right]  &=& \frac{NV}{\hbar}\sqrt{\frac{\mu_0}{\varepsilon_0}}\sum_n\frac{-2\delta \left\langle 0 \left| \Delta M_{z'} \right| n  \right\rangle \left\langle n \left| \Delta P_y' \right| 0  \right\rangle}{\left( \omega - \omega_n \right) ^2 + \delta^2},\\
\mathrm{Im}\left[ \chi_{z'z'}^{ee} \right] &=& \frac{NV}{\hbar}\frac{1}{\varepsilon_0}\sum_n\frac{-\delta \left\langle 0 \left| \Delta P_z' \right| n  \right\rangle \left\langle n \left| \Delta P_z' \right| 0  \right\rangle   }{\left( \omega - \omega_n \right) ^2 + \delta^2}, \\
\mathrm{Re}\left[ \chi_{z'z'}^{ee} \right] &=& \frac{NV}{\hbar}\frac{1}{\varepsilon_0}\sum_n\frac{\left( \omega - \omega_n \right) \left\langle 0 \left| \Delta P_z' \right| n  \right\rangle \left\langle n \left| \Delta P_z' \right| 0  \right\rangle }{\left( \omega - \omega_n \right) ^2 + \delta^2},\\
\mathrm{Im}\left[ \chi_{y'y'}^{ee} \right] &=& \frac{NV}{\hbar}\frac{1}{\varepsilon_0}\sum_n\frac{-\delta \left\langle 0 \left| \Delta P_y' \right| n  \right\rangle \left\langle n \left| \Delta P_y' \right| 0  \right\rangle   }{\left( \omega - \omega_n \right) ^2 + \delta^2}, \\
\mathrm{Re}\left[ \chi_{y'y'}^{ee} \right] &=& \frac{NV}{\hbar}\frac{1}{\varepsilon_0}\sum_n\frac{\left( \omega - \omega_n \right) \left\langle 0 \left| \Delta P_y' \right| n  \right\rangle \left\langle n \left| \Delta P_y' \right| 0  \right\rangle }{\left( \omega - \omega_n \right) ^2 + \delta^2},\\
\mathrm{Im}\left[ \chi_{y'y'}^{mm} \right] &=& \frac{NV}{\hbar}\mu_0\sum_n\frac{-\delta \left\langle 0 \left| \Delta M_{y'} \right| n  \right\rangle \left\langle n \left| \Delta M_{y'} \right| 0  \right\rangle}{\left( \omega - \omega_n \right) ^2 + \delta^2}, \\
\mathrm{Re}\left[ \chi_{y'y'}^{mm} \right] &=& \frac{NV}{\hbar}\mu_0\sum_n\frac{\left( \omega - \omega_n \right)  \left\langle 0 \left| \Delta M_{y'} \right| n  \right\rangle \left\langle n \left| \Delta M_{y'} \right| 0  \right\rangle }{\left( \omega - \omega_n \right) ^2 + \delta^2},\\ 
\mathrm{Im}\left[ \chi_{z'z'}^{mm} \right] &=& \frac{NV}{\hbar}\mu_0\sum_n\frac{-\delta \left\langle 0 \left| \Delta M_{z'} \right| n  \right\rangle \left\langle n \left| \Delta M_{z'} \right| 0  \right\rangle}{\left( \omega - \omega_n \right) ^2 + \delta^2}, \\
\mathrm{Re}\left[ \chi_{z'z'}^{mm} \right] &=& \frac{NV}{\hbar}\mu_0\sum_n\frac{\left( \omega - \omega_n \right)  \left\langle 0 \left| \Delta M_{z'} \right| n  \right\rangle \left\langle n \left| \Delta M_{z'} \right| 0  \right\rangle }{\left( \omega - \omega_n \right) ^2 + \delta^2}.
\label{eq:chis}
\end{eqnarray}
\begin{figure}[H]
	\centering
	\includegraphics[width=0.8\linewidth]{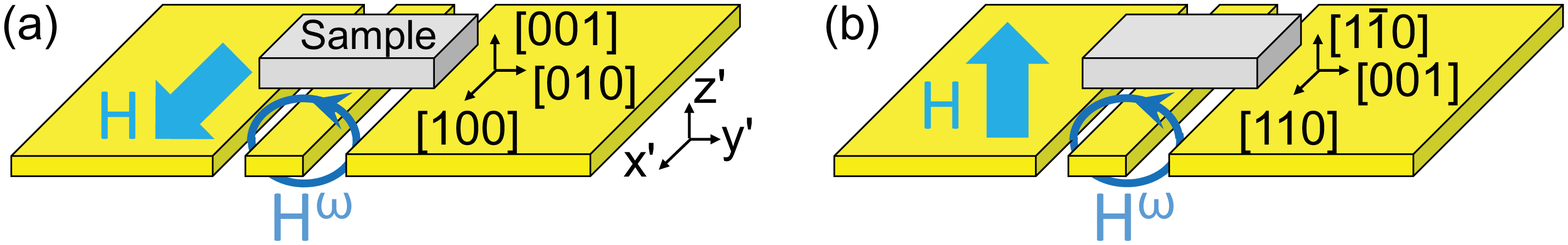}
	\caption{
		(a),(b) Experimental setups for the microwave non-reciprocity measurements (a) for $\textbf{H}\parallel[100]$, $\textbf{H}^\omega \perp [100]$ and (b) for $\textbf{H}\parallel[1\bar{1}0]$, $\textbf{H}^\omega \perp [110]$. 
	}
	\label{fig:geometries}
\end{figure}

The matrix elements of $\Delta \textbf{M}$ and $\Delta \textbf{P}$ are
\begin{eqnarray}
\left\langle 0 \left| \Delta M_{y'} \right| \alpha  \right\rangle &=& \left\langle \alpha \left| \Delta M_{y'} \right| 0  \right\rangle = 0, \\
\left\langle 0 \left| \Delta M_{y'} \right| \beta  \right\rangle &=& \left\langle \beta \left| \Delta M_{y'} \right| 0  \right\rangle = \frac{g\mu_\mathrm{B}}{V}\sqrt{\frac{S}{N}}\cos\theta \left( \cosh\phi_{1} + \sinh\phi_{1} \right), \\
\left\langle 0 \left| \Delta M_{z'} \right| \alpha  \right\rangle &=& \left\langle \alpha \left| \Delta M_{z'} \right| 0  \right\rangle = 0, \\
\left\langle 0 \left| \Delta M_{z'} \right| \beta  \right\rangle &=& -\left\langle \beta \left| \Delta M_{z'} \right| 0  \right\rangle = i\frac{g\mu_\mathrm{B}}{V}\sqrt{\frac{S}{N}}\left( \cosh\phi_{1} - \sinh\phi_{1} \right) , 
\label{eq:1DM}
\end{eqnarray}
\begin{eqnarray}
\left\langle 0 \left| \Delta P_{y'} \right| \alpha  \right\rangle & = &\left\langle \alpha \left| \Delta P_{y'} \right| 0  \right\rangle  =  0, \\
\left\langle 0 \left| \Delta P_{y'} \right| \beta  \right\rangle & = & -\left\langle \beta \left| \Delta P_{y'} \right| 0  \right\rangle = i \frac{8dflS\lambda}{V}\sqrt{\frac{S}{N}} \cos\theta \left( \cosh\phi_{1} - \sinh\phi_{1} \right), \\
\left\langle 0 \left| \Delta P_{z'} \right| \alpha  \right\rangle & = & \left\langle \alpha \left| \Delta P_{z'} \right| 0  \right\rangle  =  0, \\
\left\langle 0 \left| \Delta P_{z'} \right| \beta  \right\rangle & = & \left\langle \beta \left| \Delta P_{z'} \right| 0  \right\rangle = \frac{8dflS\lambda}{V}\sqrt{\frac{S}{N}} (2\cos^2\theta -1) \left( \cosh\phi_{1} + \sinh\phi_{1} \right), \\
\label{eq:1DP}
\end{eqnarray}
where $\left| \alpha  \right\rangle = \alpha^\dagger \left| 0 \right>$ and $\left| \beta  \right\rangle = \beta^\dagger \left| 0 \right>$. At $\omega = \omega_{2}$ $ \left( n = \alpha \right)$, the microwave absorption is zero. The relative microwave non-reciprocity at $\omega = \omega_1$ $\left( n = \beta \right)$ is
\begin{eqnarray}
\frac{\Delta\alpha}{\bar{\alpha}} &=& \frac{16dflS^2\lambda g\mu_\mathrm{B}}{V\hbar\delta} \sqrt{\frac{\mu_0}{\varepsilon_0}} \left\lbrace  \cos\theta\left(2\cos^2\theta - 1 \right)\left( \cosh\phi_1 + \sinh\phi_1 \right) ^2  - \cos\theta\left( \cosh\phi_1 - \sinh\phi_1 \right) ^2 \right\rbrace \nonumber\\
&\times& \sqrt{2}\left\lbrace \frac{Y_1}{\sqrt{X_1 + \sqrt{X_1^2 + Y_1^2}}} + \frac{Y_2}{\sqrt{X_2 + \sqrt{X_2^2 + Y_2^2}}} \right\rbrace^{-1}, 
\label{eq:Da/a}
\end{eqnarray}
where
\begin{eqnarray}
X_1 &=& \varepsilon_\infty - \left(\frac{8dflS^2g\mu_\mathrm{B}\lambda }{\hbar V\delta}\right)^2 \frac{\mu_0}{\varepsilon_0} \left( 2\cos^2\theta - 1 \right) ^2 \cos^2\theta \left( \cosh\phi_1 + \sinh\phi_1 \right) ^4,  \\
X_2 &=& \varepsilon_\infty - \left(\frac{8dflS^2g\mu_\mathrm{B}\lambda}{\hbar V\delta}\right)^2 \frac{\mu_0}{\varepsilon_0} \cos^2\theta \left( \cosh\phi_1 - \sinh\phi_1 \right)^4,  \\
Y_1 &=& \frac{\left(8dflS^{\frac{3}{2}}\lambda\right)^2}{\hbar V\varepsilon_0\delta} \left( 2\cos^2\theta - 1 \right) ^2 \left( \cosh\phi_1 + \sinh\phi_1 \right) ^2  + \varepsilon_\infty \frac{\mu_0g\mu_\mathrm{B}^2S}{\hbar V\delta} \cos^2\theta \left( \cosh\phi_1 + \sinh\phi_1 \right) ^2, \nonumber\\
&&\\
Y_2 &=& \frac{\left(8dflS^{\frac{3}{2}}\lambda\right)^2}{\hbar V\varepsilon_0\delta} \cos^2\theta \left( \cosh\phi_1 - \sinh\phi_1 \right)^2 + \varepsilon_\infty \frac{\mu_0g\mu_\mathrm{B}^2S}{\hbar V\delta} \left( \cosh\phi_1 - \sinh\phi_1 \right) ^2.
\label{eq:Da/a2}
\end{eqnarray}
The magnetic field dependence at $\omega_1$ is plotted in Fig. S4(a). Here the value of $\delta$ was estimated as $\delta =$ 1.4 GHz by the comparison of measured and calculated absorption spectra.
\begin{figure}[H]
	\centering
	\includegraphics[width=1\linewidth]{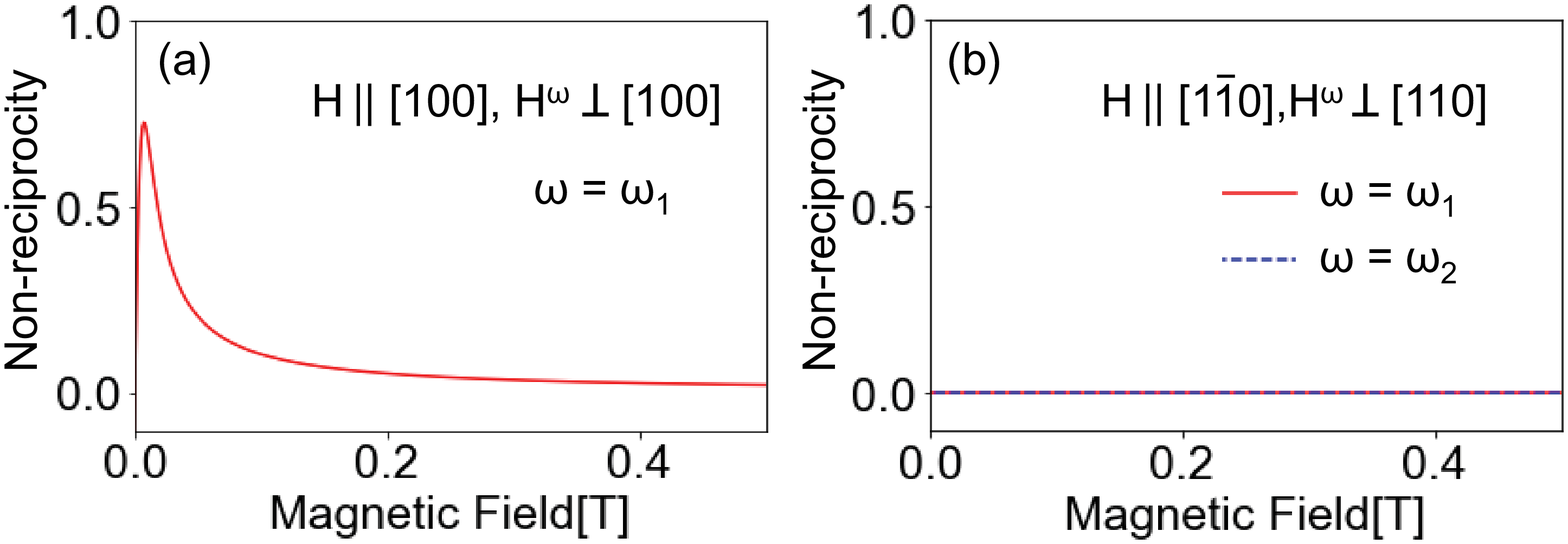}
	\caption{
		(a),(b) Relative microwave non-reciprocity $\Delta \alpha/2\bar{\alpha}=\left( \Delta S_{12} - \Delta S_{21} \right) /\left( \Delta S_{12} + \Delta S_{21} \right)$ (a) at $\omega = \omega_{1}$ for $\textbf{H}\parallel[100]$, $\textbf{H}^\omega \perp [100]$ and (b) for $\textbf{H}\parallel[1\bar{1}0]$, $\textbf{H}^\omega \perp [110]$ at $\omega = \omega_{1}$ (solid line) and $\omega = \omega_{2}$ (dashed line).
	}
	\label{fig:Dalpha}
\end{figure}

\subsection{Microwave non-reciprocity for $\textbf{H}\parallel[1\bar{1}0]$, $\textbf{H}^\omega \perp [110]$}
For $\textbf{H}\parallel[1\bar{1}0]$ and $\textbf{H}^\omega \perp [110]$, the matrix elements of $\Delta \textbf{M}$ and $\Delta \textbf{P}$ are
\begin{eqnarray}
\left\langle 0 \left| \Delta M_{y'} \right| \alpha  \right\rangle &=& \left\langle \alpha \left| \Delta M_{y'} \right| 0  \right\rangle = 0, \\
\left\langle 0 \left| \Delta M_{y'} \right| \beta  \right\rangle &=& -\left\langle \beta \left| \Delta M_{y'} \right| 0  \right\rangle = i\frac{g\mu_\mathrm{B}}{V}\sqrt{\frac{S}{N}}\left( \cosh\phi_{1} - \sinh\phi_{1} \right),\\
\left\langle 0 \left| \Delta M_{z'} \right| \alpha  \right\rangle &=& \left\langle \alpha \left| \Delta M_{z'} \right| 0  \right\rangle = \frac{g\mu_\mathrm{B}}{V}\sqrt{\frac{S}{N}}\sin\theta\left( \cosh\phi_{2} - \sinh\phi_{2} \right),\\
\left\langle 0 \left| \Delta M_{z'} \right| \beta  \right\rangle &=& \left\langle \beta \left| \Delta M_{z'} \right| 0  \right\rangle = 0,
\label{eq:2DM}
\end{eqnarray}
\begin{eqnarray}
\left\langle 0 \left| \Delta P_{y'} \right| \alpha  \right\rangle & = &\left\langle \alpha \left| \Delta P_{y'} \right| 0  \right\rangle  =  \frac{16dflS\lambda}{V}\sqrt{\frac{S}{N}}\sin\theta\cos\theta\left( \cosh\phi_{2} - \sinh\phi_{2} \right), \\
\left\langle 0 \left| \Delta P_{y'} \right| \beta  \right\rangle & = & \left\langle \beta \left| \Delta P_{y'} \right| 0  \right\rangle = 0, \\
\left\langle 0 \left| \Delta P_{z'} \right| \alpha  \right\rangle & = & \left\langle \alpha \left| \Delta P_{z'} \right| 0  \right\rangle = 0, \\
\left\langle 0 \left| \Delta P_{z'} \right| \beta  \right\rangle & = &
-\left\langle \beta \left| \Delta P_{z'} \right| 0  \right\rangle  =  i\frac{8dflS\lambda}{V}\sqrt{\frac{S}{N}}\cos\theta \left( \cosh\phi_{1} - \sinh\phi_{1} \right).
\label{eq:2DP}
\end{eqnarray}
The relative microwave non-reciprocity at $\omega = \omega_1$ $\left( n = \beta \right)$ is
\begin{eqnarray}
\frac{\Delta\alpha}{\bar{\alpha}} &=& \frac{16dflS^2\lambda g\mu_\mathrm{B}}{V\hbar} \sqrt{\frac{\mu_0}{\varepsilon_0}} \cos\theta  \left\lbrace \frac{\delta}{(\omega_2 - \omega_1)^2 + \delta^2} \left( \cosh\phi_2 - \sinh\phi_2 \right) ^2  + \frac{2\sin^2\theta}{\delta}\left( \cosh\phi_1 - \sinh\phi_1 \right) ^2 \right\rbrace \nonumber\\
&\times& \sqrt{2}\left\lbrace \frac{Y_1}{\sqrt{X_1 + \sqrt{X_1^2 + Y_1^2}}} + \frac{Y_2}{\sqrt{X_2 + \sqrt{X_2^2 + Y_2^2}}} \right\rbrace^{-1},  
\label{eq:Da/a02}
\end{eqnarray}
where
\begin{eqnarray}
X_1 &=& \varepsilon_\infty - \frac{\mu_0}{\varepsilon_0}\left( \frac{8dflS^2\lambda g\mu_\mathrm{B}}{\hbar V} \right) ^2 \frac{1}{\delta^2} \cos^2\theta \left( \cosh \phi_1 - \sinh \phi_1 \right) ^4 ,\\
X_2 &=&  \varepsilon_\infty +  \varepsilon_\infty\frac{\left(g\mu_\mathrm{B} \right)^2 S\mu_0}{\hbar V } \frac{\omega_2 - \omega_1}{(\omega_2 - \omega_1)^2 + \delta^2} \sin^2\theta \left( \cosh\phi_2 - \sinh\phi_2 \right)^2 \nonumber\\
&& + \frac{\left(16dfl\lambda \right)^2 S^3}{\hbar V \varepsilon_0} \frac{\omega_1 - \omega_2}{(\omega_1 - \omega_2)^2 + \delta^2} \cos^2\theta \left( \cosh\phi_2 - \sinh\phi_2 \right)^2 \nonumber\\
&&- \frac{\mu_0}{\varepsilon_0}\left( \frac{16dflS^2\lambda g\mu_\mathrm{B}}{\hbar V} \right) ^2 \left( \frac{\delta}{(\omega_1 - \omega_2)^2 + \delta^2} \right) ^2 \sin^4\theta\cos^2\theta \left( \cosh \phi_2 - \sinh \phi_2 \right) ^4,  \nonumber\\
&&\\
Y_1 &=& \frac{\left(8dfl\lambda\right)^2S^3}{\hbar V\varepsilon_0} \frac{\cos^2\theta}{\delta}  \left( \cosh\phi_1 - \sinh\phi_1 \right) ^2  \nonumber\\
&&+ \varepsilon_\infty \frac{\mu_0g^2\mu_\mathrm{B}^2S}{\hbar V} \frac{1}{\delta} \left( \cosh\phi_1 - \sinh\phi_1 \right) ^2,\\
Y_2 &=& \varepsilon_\infty\frac{(g\mu_\mathrm{B})^2\mu_0S}{\hbar V}\frac{\delta \sin^2\theta}{(\omega_1 - \omega_2)^2 + \delta^2} \left( \cosh\phi_2 - \sinh\phi_2 \right) ^2\nonumber\\
&&+ \frac{S^3(16dfl\lambda)^2}{\hbar V \varepsilon_0\delta}\frac{\delta \sin^2\theta\cos^2\theta}{(\omega_1 - \omega_2)^2 + \delta^2} \left( \cosh\phi_2 - \sinh\phi_2 \right) ^2 \nonumber\\
&&+ 2\frac{\mu_0}{\varepsilon_0}\left( \frac{g\mu_\mathrm{B} 8dflS^2\lambda}{\hbar V} \right) ^2 \cos^2\theta \frac{\delta (\omega_1 - \omega_2)}{(\omega_1 - \omega_2)^2 + \delta^2} \left( \cosh\phi_1 - \sinh\phi_1 \right) ^4.
\label{eq:Da/a022}
\end{eqnarray}
The relative microwave non-reciprocity at $\omega = \omega_2$ $\left( n = \alpha \right)$ is
\begin{eqnarray}
\frac{\Delta\alpha}{\bar{\alpha}} &=& \frac{16dflS^2\lambda g\mu_\mathrm{B}}{V\hbar} \sqrt{\frac{\mu_0}{\varepsilon_0}} \cos\theta  \left\lbrace \frac{2\sin^2\theta}{\delta}\left( \cosh\phi_2 - \sinh\phi_2 \right) ^2  + \frac{\delta}{(\omega_2 - \omega_1)^2 + \delta^2}\left( \cosh\phi_1 - \sinh\phi_1 \right) ^2 \right\rbrace \nonumber\\
&\times& \sqrt{2}\left\lbrace \frac{Y_1}{\sqrt{X_1 + \sqrt{X_1^2 + Y_1^2}}} + \frac{Y_2}{\sqrt{X_2 + \sqrt{X_2^2 + Y_2^2}}} \right\rbrace^{-1}, 
\label{eq:Da/a03}
\end{eqnarray}
where
\begin{eqnarray}
X_1 &=& \varepsilon_\infty + \varepsilon_\infty\frac{\left(g\mu_\mathrm{B} \right)^2 S\mu_0}{\hbar V } \frac{\omega_2 - \omega_1}{(\omega_2 - \omega_1)^2 + \delta^2} \sin^2\theta \left( \cosh\phi_2 - \sinh\phi_2 \right)^2 \nonumber\\
&& + \frac{\left(8dfl\lambda \right)^2 S^3}{\hbar V \varepsilon_0} \frac{\omega_2 - \omega_1}{(\omega_2 - \omega_1)^2 + \delta^2} \cos^2\theta \left( \cosh\phi_1 - \sinh\phi_1 \right)^2 \nonumber\\
&&- \frac{\mu_0}{\varepsilon_0}\left( \frac{8dflS^2\lambda g\mu_\mathrm{B}}{\hbar V} \right) ^2 \left( \frac{\delta}{(\omega_2 - \omega_1)^2 + \delta^2} \right) ^2 \cos^2\theta \left( \cosh \phi_1 - \sinh \phi_1 \right) ^4,  \\
X_2 &=&  \varepsilon_\infty - \frac{\mu_0}{\varepsilon_0}\left( \frac{16dflS^2\lambda g\mu_\mathrm{B}}{\hbar V} \right) ^2 \frac{\sin^4\theta\cos^2\theta}{\delta^2} \left( \cosh \phi_2 - \sinh \phi_2 \right) ^4,  \\
Y_1 &=& \frac{\left(8dfl\lambda\right)^2S^3}{\hbar V\varepsilon_0} \frac{\delta}{(\omega_2 - \omega_1)^2 + \delta^2} \cos^2\theta \left( \cosh\phi_1 - \sinh\phi_1 \right) ^2  \nonumber\\
&&+ \varepsilon_\infty \frac{\mu_0g^2\mu_\mathrm{B}^2S}{\hbar V} \frac{\delta}{(\omega_2 - \omega_1)^2 + \delta^2} \left( \cosh\phi_1 - \sinh\phi_1 \right) ^2 \nonumber\\
&&+ 2\frac{\mu_0}{\varepsilon_0}\left( \frac{g\mu_\mathrm{B} 8dflS^2\lambda}{\hbar V} \right) ^2 \cos^2\theta \frac{\delta (\omega_2 - \omega_1)}{(\omega_2 - \omega_1)^2 + \delta^2} \left( \cosh\phi_1 - \sinh\phi_1 \right) ^4,\\
Y_2 &=& \varepsilon_\infty\frac{(g\mu_\mathrm{B})^2\mu_0S}{\hbar V\delta}\sin^2\theta \left( \cosh\phi_2 - \sinh\phi_2 \right) ^2\nonumber\\
&&+ \frac{S^3(16dfl\lambda)^2}{\hbar V \varepsilon_0\delta}\sin^2\theta\cos^2\theta \left( \cosh\phi_2 - \sinh\phi_2 \right) ^2.
\label{eq:Da/a032}
\end{eqnarray}
$\Delta \alpha / 2 \bar{\alpha}$ at $\omega_1$ and $\omega_2$ are plotted in Fig. S4(b). These are quite small compared with the case of $\textbf{H}\parallel[100], \textbf{H}^\omega \perp[100]$.

\end{document}